# Morphogenesis of defects and tactoids during isotropic-nematic phase transition in self-assembled lyotropic chromonic liquid crystals

Y -K Kim[1], S V Shiyanovskii[1] and O D Lavrentovich[1]

[1]Liquid Crystal Institute and Chemical Physics Interdisciplinary Program, Kent State University, Kent, Ohio, USA

E-mail: olavrent@kent.edu

**Abstract.**
We explore the structure of nuclei and topological defects in the first-order phase transition between the nematic (N) and isotropic (I) phases in lyotropic chromonic liquid crystals (LCLCs). The LCLCs are formed by self-assembled molecular aggregates of various lengths and show a broad biphasic region. The defects emerge as a result of two mechanisms. 1) Surface anisotropy mechanism that endows each N nucleus ("tactoid") with topological defects thanks to preferential (tangential) orientation of the director at the closed I-N interface, and 2) Kibble mechanism with defects forming when differently oriented N tactoids merge with each other. Different scenarios of phase transition involve positive (N-in-I) and negative (I-in-N) tactoids with non-trivial topology of the director field and also multiply connected tactoids-in-tactoids configurations. The closed I-N interface limiting a tactoid shows a certain number of cusps; the lips of the interface on the opposite sides of the cusp make an angle different from $\pi$. The N side of each cusp contains a point defect-boojum. The number of cusps shows how many times the director becomes perpendicular to the I-N interface when one circumnavigates the closed boundary of the tactoid. We derive conservation laws that connect the number of cusps $c$ to the topological strength $m$ of defects in the N part of the simply-connected and multiply-connected tactoids. We demonstrate how the elastic anisotropy of the N phase results in non-circular shape of the disclination cores. A generalized Wulff construction is used to derive the shape of I and N tactoids as the function of I-N interfacial tension anisotropy in the approximation of frozen director field of various topological charges $m$. The complex shapes and structures of tactoids and topological defects demonstrate an important role of surface anisotropy in morphogenesis of phase transitions in liquid crystals.

# 1. Introduction

Topological defects play an important role in morphogenesis (from the Greek *morphê* shape and *genesis* creation) of phase transitions in cosmological models and in condensed matter [1-6]. In both cases, one deals with a high-temperature symmetric phase and a low-temperature phase in which the symmetry is broken. In the model of early Universe proposed by Kibble [7], topological defects such as cosmic strings form during the phase transition, when domains of the new state grow and merge. The order parameter is assumed to be uniform within each domain. When the domains with different "orientation" of space-time merge, their junctions have a certain probability of producing defects. Similar effects are expected in condensed matter systems, ranging from superfluids [6, 8] to solids [9].

One of the simplest experimental systems to explore the interplay of phase transitions and topological defects is a uniaxial nematic (N) liquid crystal (LC). In the N phase, the molecules (or their aggregates) are aligned along a direction called the director $\hat{\mathbf{n}}$ ($\hat{\mathbf{n}}^2 = 1$) with the property $\hat{\mathbf{n}} \equiv -\hat{\mathbf{n}}$ that stems from a non-polar character of ordering. In three dimensional (3D) space, he N phase allows three types of topologically stable defects: linear defects (disclinations), point defects in the bulk (hedgehogs) and point defect at the surfaces (boojums) [10].

The isotropic-nematic (I-N) transition is of the first order. Besides the transition temperature $T_{NI}$, there are two other important temperatures $T^*$ and $T^{**}$, characterizing the spinodal points: $T^* < T_{NI}$ is the limit of metastability of the I phase upon cooling and $T^{**} > T_{NI}$ is the limit of overheating the N phase. The critical size $R^*$ of N nuclei, determined by the gain in the bulk condensation energy and the loss in the surface energy of the I-N interface, is about 10 nm near $T^*$ [11]. As these nuclei grow and coalesce, they can produce topological defects at the points of junction. Chuang et al [12] and Bowick et al [13] performed the Kibble-mechanism-inspired experiments on the I-N transition and described the dynamics of ensuing defect networks. Mostly disclinations were observed, with hedgehogs appearing seldom; boojums were not recorded. A probability of forming a disclination is significant ($\sim 1/\pi$ when there are three merging domains [14]). For a hedgehog in a 3D space, as explained by Hindmarsh [15], one needs many more uncorrelated domains, which drastically reduces the probability of forming the defect.

In the analysis of defects emerging during the I-to-N phase transition, it is usually assumed that the director is roughly uniform, $\hat{\mathbf{n}}(\mathbf{r}) = \text{const}$, within each N nucleus [13, 15]. This assumption mirrors the cosmological model, in which each expanding bubble preserves spatial uniformity of the scalar field in its interior and at its surface [7, 16]. In other words, the I-N interfacial tension is considered as "isotropic", i.e. independent of director orientation at the surface. This assumption is certainly valid for small nuclei. As they grow, however, one needs to account for the anisotropy of the I-N interface, as discussed below.

Surface properties of LCs are anisotropic because the molecular interactions set up a preferred orientation of $\hat{\mathbf{n}}$ at an interface, called an "easy direction". For the I-N interface, it is convenient to introduce an "easy angle" $\bar{\alpha}$ between the normal $\hat{\mathbf{v}}$ to interface and the easy direction. Depending on the details of anisotropic molecular interactions, the easy direction might be perpendicular ($\bar{\alpha} = 0$) to the interface, conically tilted ($0 < \bar{\alpha} < \pi/2$) or tangential ($\bar{\alpha} = \pi/2$, so that $\hat{\mathbf{n}}$ can adopt any orientation in the plane of interface). To deviate the director from the easy axis to some angle $\alpha \neq \bar{\alpha}$, one needs to perform work that is in the first approximation proportional to the squared angular tilt $(\alpha - \bar{\alpha})^2$. The anisotropic potential of I-N interface is often written in the so-called Rapini-Papoular form [17] useful for analytical analysis: $\sigma = \sigma_0 \left[ 1 + w \sin^2(\alpha - \bar{\alpha}) \right]$, where $\sigma_0$ is the orientation-independent part of the surface tension and $w$ is the surface anchoring coefficient. For small angles, the Rapini-Papoular potential describes the work $\sigma_0 w (\alpha - \bar{\alpha})^2$ needed to realign the director.

The surface anchoring by itself is capable of setting up stable topological defects in the interior of each and every N domain [18, 19], in order to satisfy the theorems of Poincaré and Gauss that demand a certain number of singularities in the vector field (such as $\hat{\mathbf{n}}$) defined on surfaces with a non-zero Euler characteristic $E$ ($E=2$ for a sphere). Whether or not the director in an N nucleus of the size $R$ will follow the "easy axis", depends on the balance of the surface anchoring energy $\sim \sigma_0 w R^2$ and the elastic cost of bulk deformations $\sim KR$, where $K$ is the Frank elastic constant. The last estimate is constructed by multiplying the elastic energy density $\sim K/R^2$ by the volume $\sim R^3$ of the distorted domain. The ratio of bulk elastic constant $K$ and the surface anchoring strength $\sigma_0 w$ defines the so-called de Gennes-Kleman extrapolation length, $\xi = K / \sigma_0 w$. When the N nucleus is small, $R \ll \xi$, the director within it can be assumed to be uniform, since $KR \gg \sigma_0 w R^2$. When $R \gg \xi$ and thus $\sigma_0 w R^2 \gg KR$, the surface anchoring conditions need to be satisfied, which necessitates the existence of topological defects in each and every N drop. For example, a large spherical drop with $\bar{\alpha}=0$ must contain at least one point defect-hedgehog; a radial director field is one of the possible configurations satisfying the perpendicular surface anchoring. For typical thermotropic N materials, near the I-N transition, $K \approx 2 \text{ pN}$ and $\sigma_0 w \sim (10^{-7} - 10^{-6}) \text{ J/m}^2$ [20], thus $\xi \sim (1-10) \text{ μm}$. The assumption of a uniform director within the nuclei is thus valid when the nuclei are of a submicron size. However, each N nucleus that is larger than $\xi$ must carry topological defects as an intrinsic feature caused by surface anisotropy.

The purpose of this work is to explore experimentally the surface-anisotropy controlled morphogenesis of nuclei and topological defects during the phase transition in a LC that features a broad biphasic region of coexisting N and I phases, the so-called lyotropic chromonic liquid crystal (LCLC). Lyotropic LCs are typically formed by water dispersions of anisometric colloidal particles such as rod-like tobacco mosaic viruses (TMVs). According to Onsager, when the volume fraction of rods exceeds some critical value, the rods align parallel to each other, to maximize the translational entropy at the expense of orientational entropy [21]. The main distinctive feature of LCLCs is that their "building units" are not of a fixed shape, representing self-assembled molecular aggregates [22-25]. The range of materials which form LCLCs includes dyes, drugs [22-25], nucleotides [26] and DNA oligomers [27, 28]. Typically, the LCLC molecule is plank-like or disc-like with an aromatic flat core and peripheral polar groups. In water, the cores stack face-to-face. The aggregates, bound by weak non-covalent interactions, are polydisperse, with the length distribution that depends on concentration, temperature, ionic strength, etc. [23, 29-32]. The LCLCs show a temperature- and concentration-triggered first order I-N transition with a broad (5-15 $^0$C) coexistence region. At high temperatures, the aggregates are short and orient randomly. As the temperature decreases (or concentration increases), the aggregates elongate, multiply, and eventually some fraction of them separates from the parental I phase and forms N phase nuclei that are called tactoids [31, 33, 34]. For description of tactoids in LCs other than chromonics, see references [35-40]. The main goal of this work is to describe experimentally the surface anchoring-controlled morphogenesis of tactoids and accompanying topological defects during the I-to-N and N-to-I transitions.

The article is organized as follows. In section 2, we present a relatively simple "background" example of a thermotropic LC, formed by organic compounds within a certain range of temperatures. In these materials, the N nuclei show topological defects even when the Kibble mechanism is not relevant, i.e., the defects appear within each and every N nucleus provided its size is large enough, $R > (1-10) \text{ μm}$. In what follows, we switch to LCLCs. We limit ourselves to a two-dimensional (2D) experimental setting, described in Section 3, in which the LCLC is confined between two closely separated (a few micrometers) plates, in order to map simultaneously the details of the director field as well as the shape of N and I domains. The anisotropic nature of the I-N interface in LCLCs favors $\hat{\mathbf{n}}$ to be tangential, $\bar{\alpha} = \pi / 2$. Section 4 introduces the elements of topological description of point defects in 2D that are of two types. Point defects-disclinations of integer and semi-integer strength are located in the interior of the N domain (although they in principle might also exist at the I-N interface). Point defects-boojums of continuously defined topological charges are located at the cusps of the I-N interface. We define the cusp as a point that separates two differently tilted shoulders of the N-I interface. Section 5 describes the early stages of the I-N phase transition, in which the nucleating, growing and coalescing N tactoids produce disclinations of various strength $m = \pm 1/2$ through a process similar to the Kibble mechanism. The simplest N tactoid has two cusps and two boojums at the poles; this shape is discussed in Section 6. The scenarios of the later stages of the I-N transitions are presented in Section 7. They feature residual simply-connected I tactoids surrounded by the director fields of various strength $m = 0, \pm 1/2, \pm 1$. As the temperature is lowered, the disclinations $m = \pm 1$ split into pairs of half-integer lines. The cores of $m = \pm 1/2$ disclinations in the homogeneous N phase (formed at the end of the phase transition) show non-circular shape described in Section 8 using the idea of anisotropic elasticity of the LCLC. The reverse transition, from the homogeneous N phase with disclinations, into the I phase, is described in Section 9. The scenarios include nucleation of I tactoids either in the uniform part of the director configuration or at the cores of $m = \pm 1/2$ disclinations. The I tactoids feature cusps the number of which depends on the strength $m$ of of the surrounding director configuration $c = 2(1-m)$. The area of I tactoids grows nearly linearly with temperature, reflecting the similar temperature dependence of the volume fractions of coexisting phases. In Section 9, we also present tactoid-in-tactoid scenarios with multiple connectivity. Finally, in Section 10, we use the Wulff construction to describe the shape of topologically nontrivial N and I tactoids for elastically frozen director field with different values of $m$. The different number of cusps emerges as a natural result of surface anchoring anisotropy and the disclination-imposed requirement to have regions in which the actual director tilt $\alpha$ at the I-N interface is different from the easy axis $\bar{\alpha} = \pi/2$. The results illustrate an important role of the surface tension anisotropy in the morphogenesis of I-N phase transition.

## 2. Point defects and disclination loops in 3D nematic nuclei of thermotropic liquid crystals

Experimental exploration of morphogenesis of nuclei in the thermotropic LCs is hindered by the fact that the temperatures $T_{NI}$, $T^*$, and $T^{**}$, are all very close (one degree or so) to each other. Nevertheless, by carefully stabilizing the temperature, one can obtain and observe large N droplets that coexist with the I background, figure 1(a). We use a commercially available nematic mixture E7 (EM Industries, Inc.) comprised of cyanobiphenyl compounds. For these materials, Faetti and Palleschi [20] demonstrated experimentally that the easy angle $\bar{\alpha}$ between $\hat{\mathbf{n}}$ and the normal $\hat{\mathbf{v}}$ to the I-N interface is in the range $48°$-$65°$. The tilted easy axis implies that there is a non-zero vector projection onto the interface, $\hat{\mathbf{n}} - \hat{\mathbf{v}}(\hat{\mathbf{v}} \cdot \hat{\mathbf{n}})$. According to the Poincaré theorem, such a field must contain singularities of a total strength $\sum_i m_i$ equal the Euler characteristic of a surface, $\sum_i m_i = E$; for a sphere, $E = 2$. The strength $m_i$ of each defect is determined as a number of rotations of vector $\hat{\mathbf{n}} - \hat{\mathbf{v}}(\hat{\mathbf{v}} \cdot \hat{\mathbf{n}})$ as one circumnavigates the defect core once. The two defects on the poles of each droplet in figure 1(a), represent these singularities, called boojums, each of strength $m = 1$.

Besides the 2D topological charges $m$, one can also introduce a 3D characteristic for each boojum, defined as $A = \frac{1}{4\pi} \oiint \hat{\mathbf{n}} \cdot \left[ \frac{\partial \hat{\mathbf{n}}}{\partial \eta_1} \times \frac{\partial \hat{\mathbf{n}}}{\partial \eta_2} \right] d\eta_1 d\eta_2 = \frac{m}{2}(\hat{\mathbf{n}} \cdot \hat{\mathbf{v}} - 1) + N$ where $(\eta_1, \eta_2)$ is the pair of coordinates on a semispherical surface surrounding the boojum from the N side and $N$ is an integer [18]. In the last expression, the director field is treated as a vector and $\hat{\mathbf{v}}$ is assumed to be directed outward the N domain. Note that the definition of $A$ would be ambiguous if $\hat{\mathbf{n}}$ were treated as a director with the states $\hat{\mathbf{n}}$ and $-\hat{\mathbf{n}}$ being equivalent: Replacing $\hat{\mathbf{n}}$ with $-\hat{\mathbf{n}}$ in the definition of $A$ reverses the sign of $A$. If there are no disclination lines in the interior of the N phase, the ambiguity is easily removed by regarding $\hat{\mathbf{n}}$ as a vector rather than a director [18]. In our case, figure 1(a-c), the disclination loop near the equatorial plane of the droplets divides the surface into two parts with opposite signs of the scalar product $\hat{\mathbf{n}} \cdot \hat{\mathbf{v}}$, $S_+$ with $\hat{\mathbf{n}} \cdot \hat{\mathbf{v}} > 0$ and $S_-$ with $\hat{\mathbf{n}} \cdot \hat{\mathbf{v}} < 0$. The conservation laws connecting the 2D and 3D characteristics of the boojums in the presence of equatorial disclination has been derived in Ref. [18] as $\sum_i N_i = \frac{1}{2} \sum_i m_i = E/2 = 1$ that stem from the Gauss theorem. In the case shown in Figure 1(a-c), $A_1 = \sin^2(\bar{\alpha}/2), N_1 = 1, m_1 = 1$ and $A_2 = \sin^2(\bar{\alpha}/2), N_2 = 0, m_2 = 1$, which obeys the conservation laws above.

The surface-anchoring and topology dictated scenario of defect formation is applicable to any N nucleus that is larger than the de Gennes-Kleman length, so that it is energetically preferable to satisfy the surface anchoring conditions at the expense of the elastic deformations associated with defects. The phenomenon is not restricted to the I-N phase transition. Similar defect-rich textures are observed in equilibrium nematic droplets with a fixed size, dispersed in an immiscible isotropic fluid, such as glycerin [18], figure 1(b). The concrete set of defects depends on the easy angle $\bar{\alpha}$ that can be controlled in experiments [18]. If $\bar{\alpha}$ varies from some nonzero value to $\bar{\alpha} = 0$, then the boojums shown in Fig.1 should disappear. The disclination loop shrinks into a point defect-hedgehog at the surface (which reduces the elastic energy). The ensuing point defect can leave the surface when $\bar{\alpha} = 0$ and go into the center of drop, thus establishing either a 3D radial structure of $\hat{\mathbf{n}}$ or a more complex structure, depending on the elastic anisotropy of the material [41]. If the easy angle changes towards its maximum value $\bar{\alpha} = \pi/2$, then the disclination loop seen in Fig.1b gradually disappears, and the N drop features only two boojums that are sufficient to satisfy the tangential boundary conditions [18].

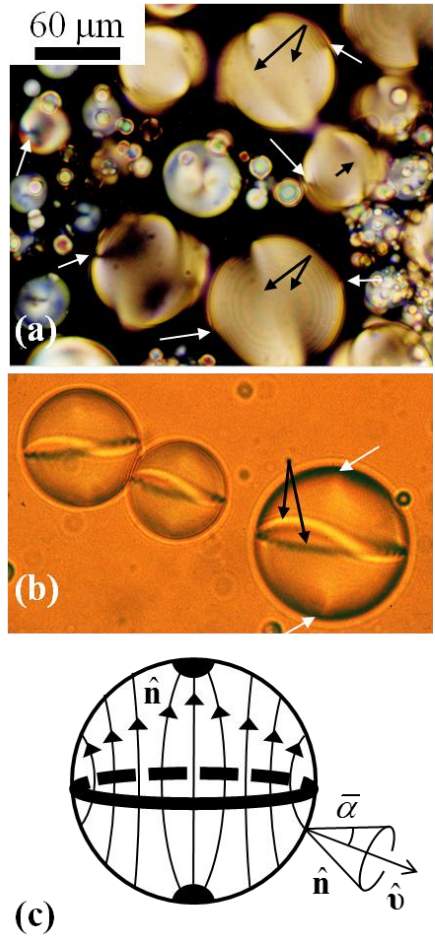

**Figure 1.** (a) Nuclei of the thermotropic nematic mixture E7 (EM Industries) emerging from the isotropic phase as viewed in a polarizing microscope with crossed polarizers; the I phase appear black; (b) spherical droplets of a thermotropic nematic n-butoxyphenyl ester of nonyloxybenzoic acid dispersed in a glycerin-lecithin matrix, viewed in a polarizing microscope with a single polarizer (no analyzer). In both systems, each N droplet shows two surface point defects-boojums (some are marked by white arrows) and disclination loops (black arrows). The topological defects occur as a result of balance of surface anchoring and elastic distortion energy; (c) principal scheme of director distortions [18, 19].

The shape of the thermotropic N droplets dispersed either in their own melt, figure 1(a), or in a foreign isotropic fluid, figure 1(b), is nearly spherical. The reason is the high surface tension of the I-N interface and a relatively weak surface anchoring: $\sigma_o \sim 10^{-5}$ J/m$^2$, while $w\sigma_o \sim \left(10^{-6}-10^{-7}\right)$ J/m$^2$ [42, 43], so that $w \sim 0.1-0.01$. The surface energy $F_s \propto \sigma_0 R^2 \sim 10^{-17}$ J of a thermotropic N droplet of a radius $R \sim 1\,\mu\text{m}$ or larger overweighs the elastic energy of internal distortions $F_e \propto K\left(\nabla \hat{\mathbf{n}}\right)^2 R^3 \sim KR \sim 2\times 10^{-18}$ J.

The interplay of surface and bulk effects during phase transitions in LCs is enriched when the materials are of lyotropic type. If the lyotropic N phase is formed by building units of a size in the range 10-100 nm, the interfacial surface energy is expected to be weaker than in the thermotropic case; experimentally, $\sigma_0 \sim \left(10^{-7}-10^{-5}\right)$ J/m$^2$ [44, 45]. The theoretical expectation [46-50] for a system of very long rigid rods is that $\sigma_0 = \chi \frac{k_B T}{LD}$, where $k_B$ is the Boltzmann constant, $T$ is the absolute temperature, $L$ and $D$ are the length and diameter of the rods, respectively, $\chi$ is the numerical coefficient estimated by different models to be in the range 0.18-0.34. Smallness of $\sigma_0$ caused by relatively large $L$ and $D$ suggests that the N droplets might not be able to maintain a spherical shape if the surface anchoring requires $\hat{\mathbf{n}}$ to be distorted in the interior. And indeed, the pioneering observations by Bernal and Fankuchen [51] of the I-N phase transition in lyotropic N formed by tobacco mosaic virus dispersed in water, revealed that the N droplets with tangential director orientation are of a peculiar elongated shape with two cusped ends [51]. These shapes were called tactoids [51, 52]; for more studies, see [35-39, 53-55].

The surface anchoring anisotropy in lyotropic LCs might be more pronounced than in their thermotropic counterparts. The experimental estimates range from $w \approx 4$ [56] for water dispersions of carbon nanotubes to $w \sim 10-100$ ($w\sigma_0 \sim \left(0.5-5\right)\times 10^{-5}$ J/m$^2$ and $\sigma_0 \sim 5\times 10^{-7}$ J/m$^2$) for vanadium pentoxide dispersions [35, 36]. Theoretical consideration [57] for lyotropic systems of wormlike chains predicts a simple Rapini-Papoular type angular dependency of the surface tension, $\sigma(\alpha) = \sigma_0\left(1+w\cos^2\alpha\right)$, with $w \sim 0.5-1$ (depending on flexibility of chains).

On the other hand, the Frank elastic constants in the lyotropic LCs are nearly of the same order as those in the thermotropic LCs [32]. It is thus expected that the structure of nuclei in the I-N phase transitions would be highly nontrivial, both in terms of their shape and the interior director structure, as the representative energies $\sigma_0 R^2$, $\sigma_0 w R^2$, and $KR$ might vary in a much wider range than in the thermotropic LCs. Furthermore, in thermotropic systems with a relatively weak surface anisotropy, the Rapini-Papoular surface potential or even a simpler quadratic dependence $\sigma(\alpha) \propto (\alpha - \bar{\alpha})^2$ is often sufficient to describe the anchoring phenomena. It might not be true in the case of a lyotropic LC with a large $w$; observation of the first-order anchoring transition of LCLCs in contact with solid substrates suggests that the surface potential should be different from the Rapini-Papoular form [58].

## 3. General properties of LCLCs and experimental techniques

We explore water solutions of disodium cromoglycate (DSCG) (Sigma-Aldrich, purity $\geq 95\%$), one of the first studied LCLCs [22, 24], figure 2(a). In water, the DSCG molecules stack on top of each other face-to-face (the so-called H-aggregation) to minimize the areas of unfavorable contact with water [31, 33, 59]. The stacking distance is (0.33 - 0.34) nm [31, 59]. The stacking distance makes LCLC aggregates similar to the double-strand B-DNA molecules. The important difference is that in LCLCs, there are no chemical bonds to fix the length of aggregates.

We used DSCG concentration $c = 16 \, wt\% \, (= 0.37 \, \text{mol/kg})$. The relevant portion of phase diagram is shown in figure 2(c); note a broad biphasic region in which the N and I phases co-exist. Figure 2(d) shows how the area of the N phase changes with temperature in the biphasic region. Although the trend is close to the linear one, the curves for heating and cooling are somewhat different, since the director pattern in the sample is different upon cooling and heating.

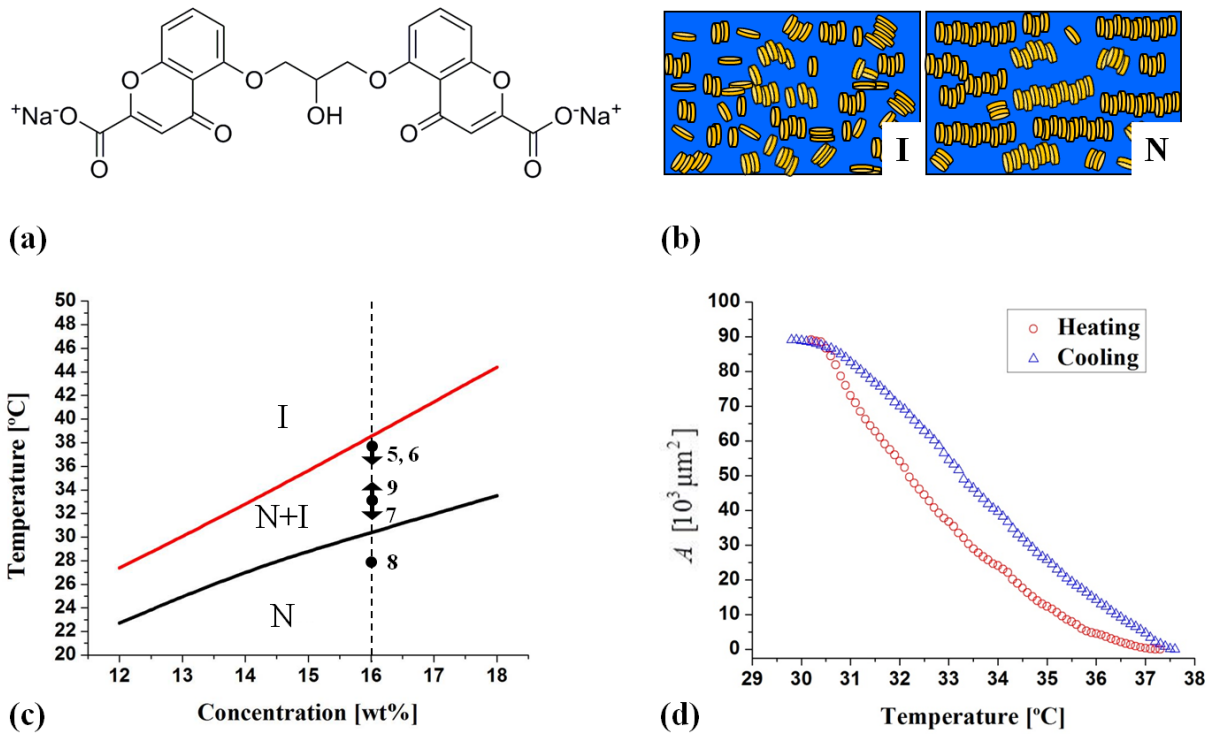

**Figure 2.** (a) Molecular structure of DSCG. (b) Schematic structure of chromonic aggregates in I and N phases. (c) Phase diagram of DSCG dispersed in water as the function of temperature and DSCG concentration. The numbers correspond to the chapters that describe the scenarios of phase transition; the dots indicate the approximate location of the system on the phase diagram and the arrows indicate whether the system is cooled down or heated up; the dot with the number 8 and no arrow corresponds to the discussion of disclination cores in Chapter 8. (d) Temperature dependence of the area occupied by the N phase in biphasic region, for heating and cooling; the rate of temperature change is $\pm 1^\circ\text{C/min}$. Note the difference in the two curves, associated with the different type of director distortions in the system.

From the experimental point of view, it is important to explore a pseudo-2D geometry, to mitigate complications associated with the effects such as depth-dependent $\hat{\mathbf{n}}$. 2D geometry also allows one to apply a quantitative methods of optical microscopy such as mapping of the optical retardance and orientation of $\hat{\mathbf{n}}$ [60]. We remind the reader that in the regular optical polarizing microscopy, the image is determined by the integral over the (generally distorted) configuration of the optic axis along the pathway of light propagation. If the sample is thin, the undesirable director distortions along the light propagation direction are suppressed and the image represents a 2D pattern of the director field that can be reconstructed by using a microscope with LC PolScope universal compensator [61].

The samples were prepared as thin slabs of thickness $d = (1-5)\,\mu\text{m}$ between two glass plates, spin-coated with un-rubbed polyimide SE-7511 (Nissan, Inc.) for tangential anchoring of the DSCG director. The cell thickness was set by glass spheres mixed with UV epoxy (Norland Optical Adhesive 65, Norland Products, Inc.) applied at the periphery of cells to seal them and to prevent evaporation of water. The temperature was controlled with the Linkam controller TMS94 and hot stage LTS350 (Linkam Scientific Instruments) with precision of $0.01^{\circ}\text{C}$. In both cooling and heating, the temperature rate was typically $0.1^{\circ}\text{C}/\min$; we waited until the expansion of N or I tactoids would stop before changing the temperature again.

The textures were examined by a polarizing microscope (Nikon E600) equipped with Cambridge Research Incorporation (CRI) Abrio LC-PolScope package. The LC PolScope uses a monochromatic illumination at 546 nm and maps optical retardance $\Gamma(x,y)$ and orientation of the slow axis in the sample [60]. The maximum measured optical retardance in the commercial CRI Abrio LC-PolScope package is listed as 273 nm, but we found by testing wedge samples with a variable retardance that the device does not produce accurate measurement for any value of $\Gamma(x,y)$ larger than about 240 nm. We thus restricted the thickness $d$ of our cells to set $\Gamma(x,y)$ in the reliable range between 0 and 240 nm. For a tangentially anchored N, $\Gamma = |n_e - n_0|d$, where $n_e$ and $n_o$ are the extraordinary and ordinary refractive indices. For DSCG, optical birefringence is negative, $n_e - n_0 \approx -0.02$ [33, 34]. The slow axis is thus perpendicular to the optic axis $\hat{\mathbf{n}}$. The PolScope was set up to map the local orientation $\hat{\mathbf{n}}(x,y)$ (rather than the slow axis), by headed nails. Note that the nail's head is an artificial feature of PolScope software that is not related to any particular property of the LCLC, as $\hat{\mathbf{n}} = -\hat{\mathbf{n}}$ and there is no director tilt at the N-substrate interface. The accuracy with which the PolScope determines the orientation of optic axis is better than 0.1 degree. The map of retardance $\Gamma(x,y)$ allow us to trace the changes in the degree of the scalar order parameter $S$, as in the first approximation [62], $|n_e - n_0| \propto S$.

The surface anchoring of DSCG at bounding plates is tangential. The director configuration can be treated as 2D, as the bounding plates suppress the out-of-plane distortions. These distortions are better suppressed in thinner samples. However, different contact angles between the N and I phases and the bounding plates might still distort $\hat{\mathbf{n}}$ in the vertical cross-section of the cell. To explore the menisci, we used fluorescent confocal polarizing microscopy (FCPM) that images vertical cross-sections of the samples [63]. The DSCG solution was doped with a water-soluble fluorescent dye Acridin Orange (Sigma-Aldrich); it concentrates predominantly in the I phase. In this particular experiment, we used thick cells of 20 μm to enlarge the menisci. Figure 3 shows that the I-N interface is nearly perpendicular to the bounding plates, with the N side forming a contact angle $\beta$ in the range around $(77 \pm 7)^0$ with the substrate. Since the contact angles are not very different from $\pi/2$, the meniscus profile should not significantly alter the shape of N and I regions that is determined from the PolScope images of thin samples. In a sample of thickness 2 μm, the meniscus effect would lead to a 0.2 μm inaccuracy in measured lateral distances, which is less than the optical resolution (about 0.5 μm).

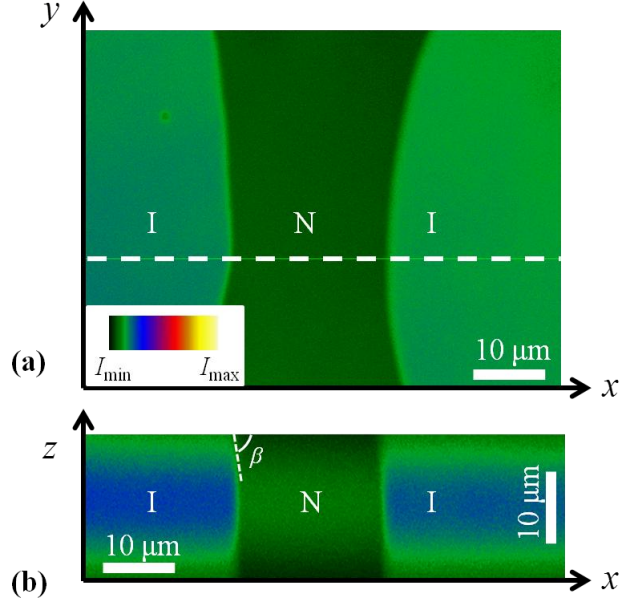

**Figure 3.** FCPM images of the N-I biphasic region in DSCG water solution doped with fluorescent dye: (a) in-plane and (b) vertical cross-section of the cell. The tilt angle of I-N interface with respect to the bounding plate is about $(77 \pm 7)^0$.

In order to estimate the surface tension $\sigma_0$ of the I-N interface, we used the pendant drop technique [44, 64]. The 16 wt% solution of DSCG was centrifuged at 4400 rpm at N-I biphasic temperature, $T = 38^\circ \text{C}$, to achieve macroscopic phase separation. The I and N phases were carefully filled into separate syringes. The corresponding densities were measured with DE45 Density Meter (Mettler Toledo) at $T = 38^\circ \text{C}$: $\rho_N = 1.08 \, \text{g/cm}^3$ and $\rho_I = 1.06 \, \text{g/cm}^3$. The centrifuged I solution was transferred into the rectangular bath and the pipette containing the N solution was inserted into the I bath kept at $38^\circ \text{C}$. The N solution was pushed out from the pipette to form a pendant drop. The interfacial tension $\sigma_0$ was determined by fitting the drop profile with the theoretical plots [44, 64]. Since the droplets were stable only within minutes, we could only estimate the order of magnitude, $\sigma_0 \approx 10^{-4} \, \text{J/m}^2$.

**4. Topological characteristics of point defects in 2D N: disclinations and boojums.**

In what follows, we operate with the 2D description of topological defects in the N domains, Fig.4. The 2D director field is parameterized as $(n_x, n_y) = [\cos\Phi(\theta), \sin\Phi(\theta)]$, where $\Phi$ is the angle between the director and a fixed axis $x$ in the $(x, y)$ plane, and $\theta$ is the polar angle of the polar coordinate system $(r, \theta)$, Fig.4(g). The order parameter space of a 2D N represents a circle $S^1/Z_2$ with two opposite points being identical to each other. The first homotopy group of $S^1/Z_2$ is nontrivial, $\pi_1(S^1/Z_2) = \{0, \pm 1/2, \pm 1, ...\}$, which implies that in the interior of the 2D domains, there might be topologically stable point disclinations of various "strength" or "topological charge" [10]. The disclination strength is introduced as an integer or semi-integer number,

$$m = \frac{1}{2\pi} \oint \left( n_x \frac{\partial n_y}{\partial \theta} - n_y \frac{\partial n_x}{\partial \theta} \right) d\theta = \frac{1}{2\pi} \left[ \Phi(\theta = 2\pi) - \Phi(\theta = 0) \right] = 0, \pm 1/2, \pm 1, \ldots. \quad (1)$$

For example, a simple radial configuration $(n_x, n_y) = (\cos\theta, \sin\theta)$ yields $m = 1$. Four main disclination types with $m = \pm 1/2, \pm 1$ are shown in figures 4(a-d); as an example of $m = 1$ disclinations, in figure 4(c), we show a circular director pattern, $(n_x, n_y) = [\cos(\theta + \pi/2), \sin(\theta + \pi/2)]$.

The second type of point defects, boojums, are observed in the experiments with DSCG as defects located at the cusps of I-N interface. The "strength" of a boojum is determined not only by the behavior of $\Phi(\theta)$ but also by the angle $\tau_N$ between the two shoulders on the opposite sides of the cusp (measured in the N phase anti-clockwise; we direct the $x$ axis along one of the shoulders):

$$m = \frac{1}{2\pi} \int_0^{\tau_N} \left( n_x \frac{\partial n_y}{\partial \theta} - n_y \frac{\partial n_x}{\partial \theta} \right) d\theta = \frac{1}{2\pi} \left[ \Phi(\theta = \tau_N) - \Phi(\theta = 0) \right]. \quad (2)$$

For a flat N-I interface, $\tau_N = \pi$ and the radial director field $(n_x, n_y) = (\cos\theta, \sin\theta)$ yields $m = 1/2$, an intuitively clear result, as such a boojum represents ½ of the radial bulk disclination $m = 1$. When the interface is not flat, $\tau_N \neq \pi$, equation (2) results in

$$m = \frac{k}{2} - \frac{\tau}{2\pi}, \quad (3)$$

where $k$ is an integer and $\tau = \pi - \tau_N$ is the angle measured between the right shoulder and the continuation of the left shoulder of the cusp. The convenience of last notation is that the sign of $\tau$ discriminates between the positive cusp (N phase protruding into the I phase, $0 < \tau_N < \pi$, $0 < \tau < \pi$), and the negative cusp (I phase protruding into the N phase, $\pi < \tau_N < 2\pi$, $-\pi < \tau < 0$).

The sign of $m$ and $k$ is determined uniquely by the comparison of the traveling direction around the defect core and the sense of director rotation [10]. Figure 4(e) shows a positive boojum with $k = 1$, while figure 4(f) shows a negative boojum with $k = -1$. This boojums have the lowest energies and therefore are observed in our experiment. A 'zero' boojum with $k = 0$ is unstable as it exists only with a cusp and can be smoothly eliminated simultaneously with the cusp, $\tau \to 0$. It is important to stress that for the tangential surface anchoring at the I-N interface (easy angle $\bar{\alpha} = \pi/2$), the distinction between the disclinations and boojums is of an energetic rather than a topological origin. The presence of cusps filled with boojums is dictated by the balance of surface and elastic energies; when such a balance yields $\tau = 0$, the boojums acquire semi-integer or integer strength and can leave the interface and move into the interior of the N domains. If the 2D picture is extended to the third dimension (along the normal to the LCLC cell), then the point disclinations would correspond to linear disclinations and the point boojums to the linear surface disclinations parallel to the third dimension axis.

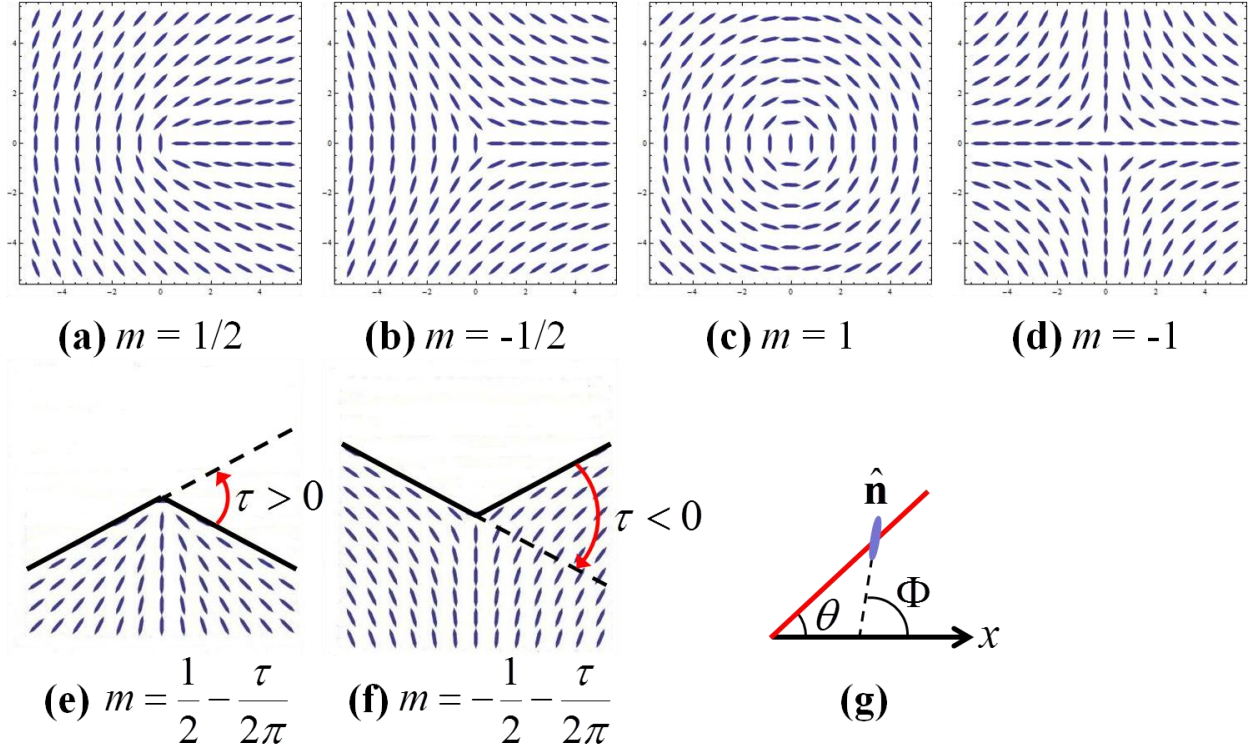

**Figure 4.** Director configurations for (a-d) disclinations $m = \pm 1/2, \pm 1$, (e) positive boojum $m = \frac{1}{2} - \frac{\tau}{2\pi}$ inside a positive cusp, (f) negative boojum $m = -\frac{1}{2} - \frac{\tau}{2\pi}$ associated with a negative cusp, (g) schematic definition of angular parameters.

### 5. Early stages of I-to-N transition: tactoids, boojums and disclinations.

Figure 5 illustrates appearance of the N phase through elongated tactoids with a two-cusped shape, or "2c tactoids", when the temperature is lowered from the homogeneous I phase into the biphasic phase, see label "5" in figure 2(c). There are two surface defects-boojums at the cusps. The I-N interface near the cusp forms an angle $0 < \tau < \pi$, figure 6(b), so that the corresponding boojum is of a strength $m = 1/2 - \tau/2\pi$, figure 4(e). The director is tangential to the N-I interface, as clear from the image of a large tactoid in figure 5(a). In the cusp regions, the retardance is reduced as compared to the interior of tactoid. The reasons are (i) non-flat meniscus, (ii) finite width of the interfacial region and decrease of the scalar order parameter in order to reduce the energy of strong director distortions, (iii) realignment of $\hat{\mathbf{n}}$ along the vertical axis, an effect similar to the "escape into the third dimension" [62].

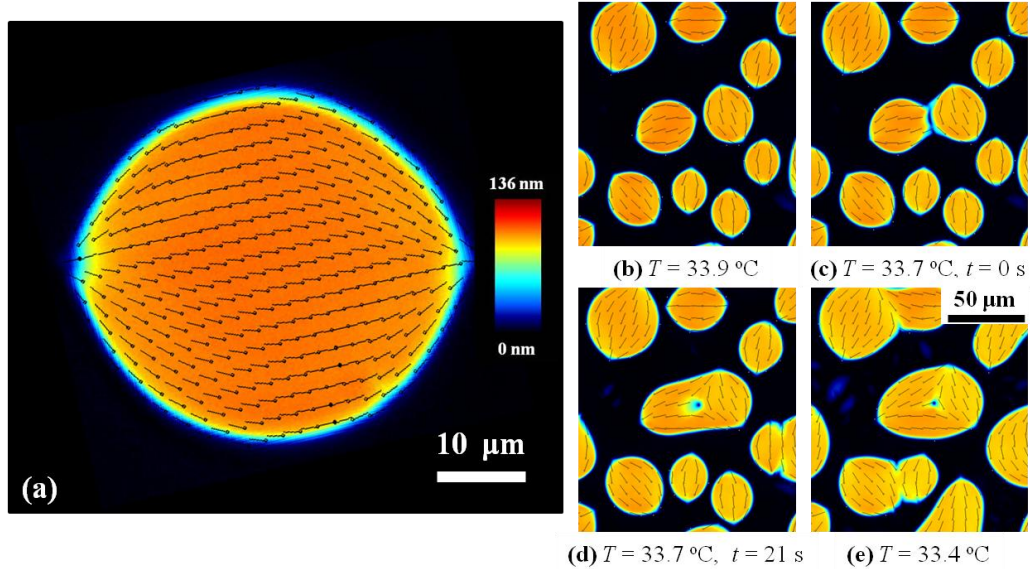

**Figure 5.** PolScope textures of nucleating N tactoids during the I-to-N phase transition in DSCG. (a) The N tactoids feature two cusps with point defects-boojums and tangential orientation of the director at the I-N interface. (b-e) The tactoids grow and merge as the temperature is reduced. Note that a merger of two differently oriented 2c tactoids in (c) produces a 3c tactoid with a disclination $m = -1/2$ in the center (d,e).

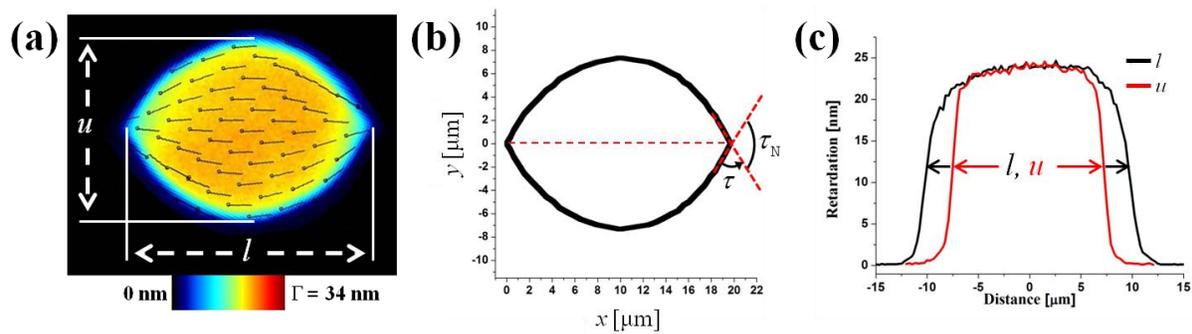

**Figure 6.** Anisotropic shape of a small 2c tactoid: (a) PolScope texture and (b) reconstructed shape; (c) phase retardance $\Gamma$ measured along the long $l$ and short $u$ axes of tactoid.

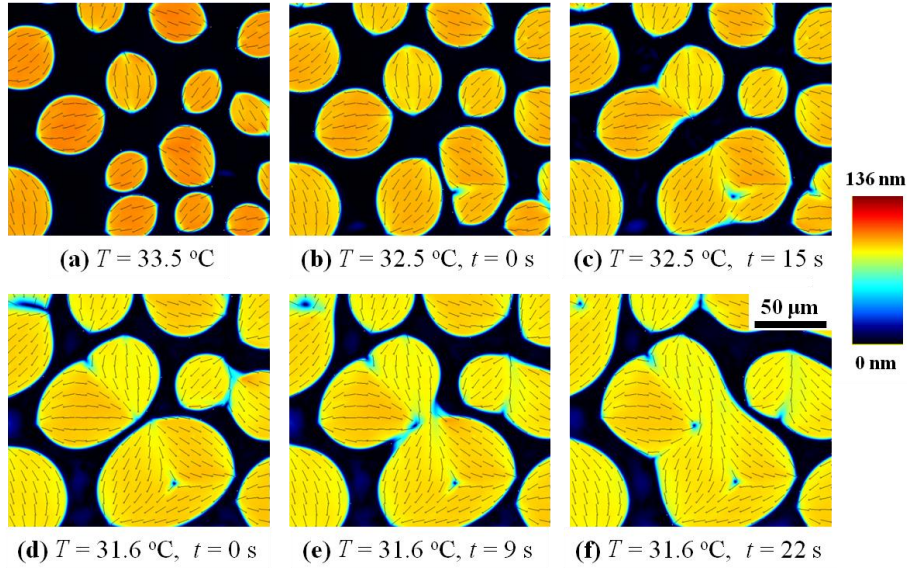

**Figure 7.** PolScope textures of N tactoids with (a) 2 cusps, (b) 4 cusps, (c) 3 cusps and one disclination $m=-1/2$, (d-f) coalescence of two large tactoids into a single tactoid with a pair of oppositely charged disclinations $m=\pm 1/2$ in the bulk, 4 positive and 2 negative cusps at the I-N interface.

As the temperature is lowered, the 2c tactoids grow and coalesce. Most of the time, the tactoids coalesce in pairs and the new tactoid eventually regains the same 2c shape. However, the merger can also produce nontrivial topological defects in the bulk, similarly to the Kibble mechanism. For example, parts (c)-(e) in figure 5 show a formation of a $m=-1/2$ interior disclination as a result of coalescence of two 2c tactoids with practically orthogonal alignment of long axes. The resulting tactoid has three cusps at the I-N interface, associated with three boojums, each of a positive strength $0<m<1/2$. The difference between the described scenario of coalescence-triggered defect formation from the classic Kibble mechanism is that each tactoid participating in the coalescence acts at the scales of (1-100) μm and larger, already contains the topological defects caused by the surface anchoring phenomenon.

Coalescence also produces "negative" cusps of protruding I phase, the N part of which is filled with a boojum of a negative strength $-1/2<m<0$, figure 4(f). Three negative cusps are clearly visible in figure 5(e) and figure 7(f). Since each N domain is topologically equivalent to a disk, for the fixed in plane tangential surface anchoring, the total director reorientation measured by circumnavigating the I-N boundary, should be equal $2\pi$ minus the angle that describes missing orientations of the I-N interface. If there are $i=1,2,...c$ boojums at the interface of strengths $m_i$ associated with the cusps of angles $\tau_i$ and $j=1,2,...n$ disclinations of charges $m_j$ in the interior, then the conservation law for the topological charges involved writes

$$\sum_i^c \left(m_i + \frac{\tau_i}{2\pi}\right) + \sum_j^n m_j = 1 \quad \text{or} \quad \frac{1}{2}\sum_i^c k_i + \sum_j^n m_j = 1 \qquad (4)$$

For a smooth round disk, $\tau_i = 0$, the last relationship reduces to $\sum_i^c m_i + \sum_j^n m_j = 1$. In our experiments only low energy boojums are observed: a positive boojum, $k=1$, inside the positive cusp, figure 4(e) and a negative boojums $k=-1$ in the negative cusp, figure 4(f). Then the relationship (4) can be rewritten in terms of the excess number of positive cusps $c^+$ over the negative ones $c^-$ that is determined by the total topological charge of disclinations:

$$c^+ - c^- = 2\left(1 - \sum_{j}^{n} m_j\right). \tag{5}$$

As an illustration, the bottom-right part of figure 7(b) shows a tactoid with 3 positive cusps and one negative cusp, and no bulk disclinations, thus $3-1=2$. When such a tactoid coalesces with a regular 2c tactoid the negative surface boojum evolves into a full bulk disclination $m=-1/2$, figure 7(c). The next merger between a disclination-free 4c tactoid and 3c tactoid with a disclination, figure 7(d-e), produces a large 6c tactoid with 4 positive cusps, 2 negative cusp, one disclination $m=-1/2$ and one disclination $m=1/2$, so that the total number of cusps and the topological charges are again satisfying equation (5), as $4-2=2(1-1/2+1/2)$, figure 7(f). Mergers of defects in the bulk and at the I-N interface is possible, but it does not change the sum total.

The long axes of different tactoids are oriented in different directions. The local alignment at a particular point $x_0, y_0$ might be easily changed when the domain coalesce and restructure their shape and the director field, as seen by comparing parts (c)-(f) of figure 7. These features indicate that there is no surface-imposed local anchoring direction in the $x, y$ plane and that the director configurations and shape of the tactoids are determined by the balance of elastic distortions energy and anisotropic surface tension of the I-N interface rather than by the in-plane anchoring of the glass substrates.

The color in PolScope textures of the tactoids in figure 7 is gradually changing from the reddish to yellow as the temperature is lowered and the area occupied by the N phase increases. It shows that the N domains at higher temperatures have a higher concentration of DSCG (and thus higher birefringence) than their counterparts at the lower temperatures [33].

## 6. Shape of positive N tactoids with two cusps.

The shape of negative and positive tactoids is determined by balance of the elastic energy of the director in the N phase and by the I-N interfacial tension energy. The problem is mathematically challenged and can be solved analytically only for certain simplified situations, see, for example, [40, 65-67] and references therein. One of those is a special case of crystals, in which the elastic bulk forces are infinitely large as compared to the anisotropic surface forces [68]. In such a case, the crystal shape is described by the classic Wulff construction based on the angular dependence of $\sigma$. The Wulff construction has been applied to the elastically "rigid" and uniform N drops with $\hat{\mathbf{n}}=\text{const}$. As discussed by Herring [69, 70], if the angular dependence $\sigma(\alpha)$ on the angle $\alpha$ between $\hat{\mathbf{n}}$ and $\hat{\mathbf{v}}$ is pronounced (but $\sigma$ remains a smooth function of $\alpha$ with continuous derivatives), the equilibrium shape should be bounded by a number of smoothly curved (not flat) surfaces which intersect in sharp edges. Herring [69] cited the observations by Zocher of nematic tactoids with two sharp cusps [52] as an experimental confirmation of Wulff construction. The only experimental case with unperturbed director, $\hat{\mathbf{n}}=\text{const}$, in the interior of tactoids was reported recently by Puech et al [56] for a water dispersion of carbon nanotubes. The inverse Wulff construction led to the estimate $w \approx 4$. Kaznacheev et al [35, 36] found that for vanadium pentoxide dispersion in water, surface anisotropy is even stronger, $w \sim 10-100$.

In our case, the director field is clearly distorted within the 2c tactoids, being tangential to the I-N interface. We are not aware of any analytical results for such geometry. Recently, van der Schoot et al [71] performed numerical simulations of the shape of positive 2c tactoids taking into account the anisotropic surface tension in the form $\sigma = a_0 + a_2 \cos 2\alpha$ and internal elasticity in one-constant approximation with the Frank modulus $K$. In our notations, $\sigma = \sigma_0\left(1 + w\cos^2\alpha\right)$, $\sigma_0 = a_0 - a_2$ and $\sigma_0 w = 2a_2$. The shapes and director textures within the tactoids were simulated in [71] for various values of $\gamma = \frac{2a_2}{K}\sqrt{\frac{A}{\pi}}$ and $\omega = \frac{2a_0}{K}\sqrt{\frac{A}{\pi}}$, where $A$ is the surface of the 2D tactoid. Unfortunately, the upper limit of $\omega$ was only 100, which limited the comparison of our experimental shapes with those simulated in [71], to very small tactoids. For a relatively small tactoid in figure 6, of a surface area $A = 200\,\mu m^2$, we measure the ratio $\varepsilon = l/u$ of the long axis (length $l$) to the short axis (width $u$) to be around $1.3 \pm 0.1$ and the cusp angle to be $\tau_N = 1.05 \pm 0.05$. The radius of curvature of the I-N interface at the cusps is less than the resolving power of our microscope, thus smaller than $0.5\,\mu m$. Comparison to the simulation results in figure 9 of [71] shows that the experimentally observed N tactoids correspond to the so-called "III regime" of [71]. In this regime, $\hat{n}$ is always parallel to the I-N interface, while the shape is modestly anisotropic, $\varepsilon < 2$, and has two pronounced cusps with boojums. Comparison with figure 8 in [71] leads to a rough estimate $\gamma/\omega \sim 0.85 \pm 0.1$, which suggests a rather strong anisotropy of the interfacial tension, on the order of $w \sim 10$. To measure $w$ in a more reliable manner, we would need to determine the surface tension more accurately than presently possible and expand the range of numerical simulations to the range of parameters characterizing the tactoids in LCLCs; this work is in progress.

## 7. Late stages of I-to-N transition: I tactoids as disclination cores

At the late stages of the I-to-N transition, when many large domains coalesce, they occasionally trap I islands around which the director rotates by $2\pi$ or $-2\pi$, which corresponds to integer strength $m = 1$ (figure 8) and $m = -1$ (figure 9) disclinations. Topologically, these are allowed in the 2D case. When the temperature is reduced and the I islands at the core shrinks, the integer disclinations always split into pairs of semi-integer disclinations, either $m = 1/2$ (figure 8), or $m = -1/2$ (figure 9). In 3D LCs, the integer disclinations do not show such a split, as $\hat{n}$ simply realigns parallel to the defect's axis ("escape into the third dimension"). It is only for samples thinner than $0.5\,\mu m$ that one can suppress the escape by a very strong tangential anchoring of $\hat{n}$ [72]. In our case of LCLC textures, the escape is apparently suppressed at higher thicknesses as the integer cores split into pairs of semi-integer disclinations. The effect indicates that the director remains mostly parallel to the $(x,y)$ plane.

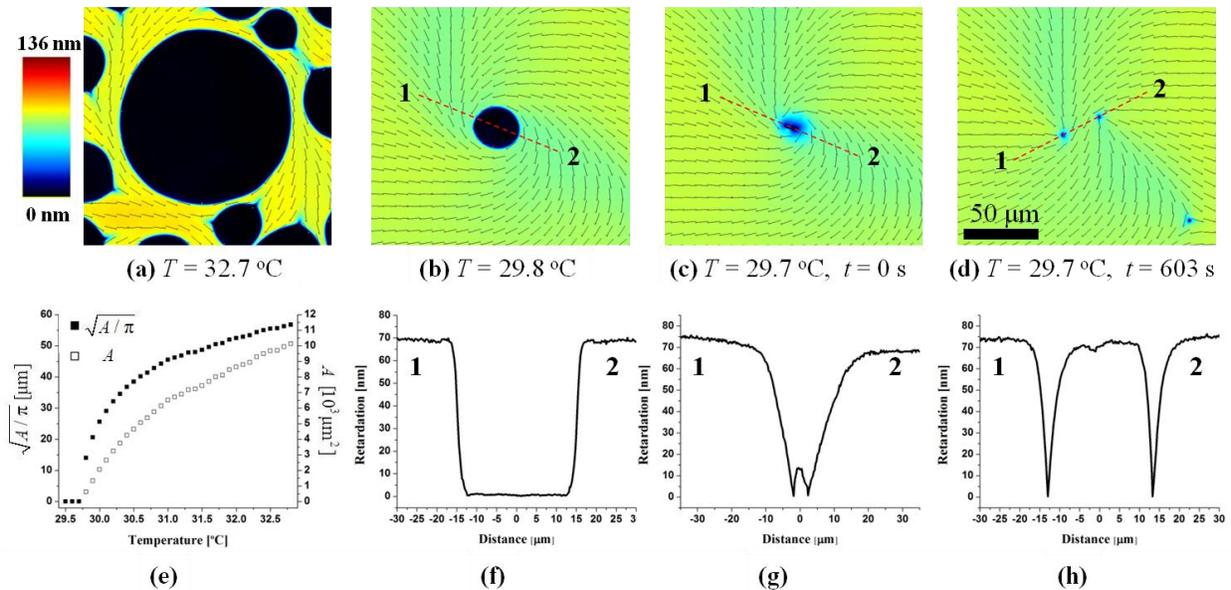

**Figure 8.** (a) PolScope textures of a round I domain associated with a $m = 1$ director field around it; (b),(c) the I tactoid shrinks as the temperature is reduced and then (d) splits into two disclinations of strength $m = 1/2$ each. (e) Temperature dependence of characteristic size and area of the I tactoid. (f,g)

Retardance profile of the I tactoid shown in parts (b,c), respectively. (h) Retardance profile of the two disclinations in part (d) separating from each other.

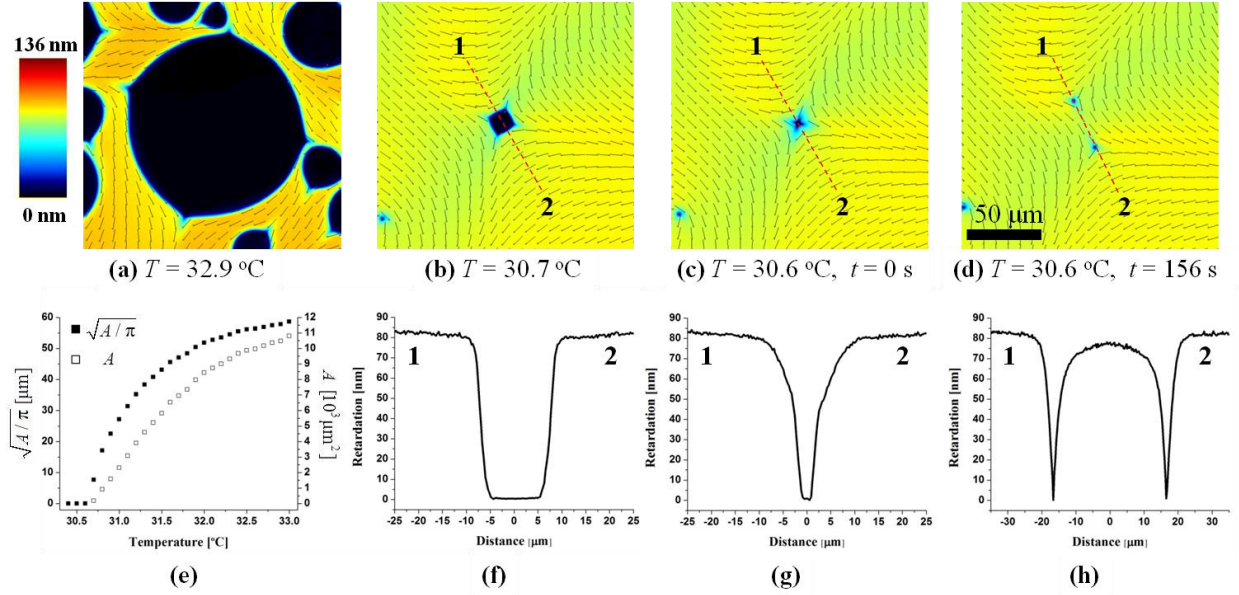

**Figure 9.** (a) PolScope textures of a 4c negative tactoid associated with a $m = -1$ director field around it; (b),(c) the I tactoid shrinks as the temperature is reduced and then (d) splits into two disclinations of strength $m = -1/2$ each. (e) Temperature dependence of characteristic size and area of the 4c negative tactoid. (f,g) Retardance profile of the 4c negative tactoid shown in parts (b,c), respectively. (h) Retardance profile of the two disclinations in part (d) separating from each other.

The elastic energy of the disclination with the director perpendicular to its axis grows as $m^2$: $f_e = \pi m^2 K \ln R/r_c + f_c$, where $K$ is the average value of the Frank splay and bend elastic constants, $R$ is the size of the system or the distance to the closest disclination, $r_c$ and $f_c$ are the radius and the energy of the core, respectively. The elastic energy $\sim m^2$ thus favors splitting of the integer disclinations into pairs of semi-integer defects. On the other hand, the interfacial energy of the N-I interface should favor the $m = 1$ structure with a single isotropic core. Consider the effect in a greater detail.

The energy of $m = 1$ disclination (per unit length, assuming $\hat{\mathbf{n}}$ is perpendicular to the axis) is comprised of the elastic energy $f_{e,1} = \pi K \ln R/r_1$ of the director, the excess energy of the isotropic core, $f_{cond,1} = (f_I - f_N)\pi r_1^2$, where $f_I$ and $f_N$ are the free energy densities of the I and N phases, respectively, and the energy of the I-N interface $f_{IN,1} = 2\pi r_1 \sigma_0$. Minimization of $f_1 = f_{e,1} + f_{cond,1} + f_{IN,1}$ with respect to $r_1$ yields $r_1 = \frac{1}{2(f_I - f_N)}\left(\sqrt{\sigma_0^2 + 2K(f_I - f_N)} - \sigma_0\right)$. For the split pair of disclinations $m = 1/2$, separated by a distance $L$, the elastic energy is $f_{e,pair} = \pi K \ln R/r_{1/2} - \frac{\pi}{2} K \ln L/2r_{1/2}$, while $f_{cond,pair} = 2(f_I - f_N)\pi r_{1/2}^2$ and $f_{IN,pair} = 2\pi r_{1/2}\sigma_0(2 + w)$. The latter estimate accounts for a change of director orientation from tangential to normal when one circumnavigates the defect core shown in figure 4(a). To estimate at which temperature the defect $m = 1$ can split into the pair of $m = 1/2$ defects, we assume that at the moment of splitting, the areas of the cores are equal, i.e. $r_{1/2} = r_1/\sqrt{2}$. Then the difference in the free energies of the two configurations can be written as

$$f_1 - f_{pair} = \frac{\pi K}{2} \ln \frac{L}{\sqrt{2}r_1} + \sqrt{2}\pi r_1 \sigma_0 \left(\sqrt{2} - 2 - w\right). \tag{6}$$

The first term favors splitting while the second term stabilizes the single-core disclination. Assuming that the defect core follows the experimentally observed behavior $\propto \sqrt{A}$, we approximate its temperature dependence as

$$r_1 = \sqrt{\frac{A}{2\pi}\left(\tanh\frac{2T - T_I - T_N}{2T_I - 2T_N} + 1\right)}, \tag{7}$$

where $T_I$ and $T_N$ are the upper and lower boundaries of the biphasic region, respectively. Then equation (6) predicts that the quantity $f_1 - f_{pair}$ can change the sign from negative to positive as the temperature is lowered. An example of $f_1 - f_{pair}$ vs. $T$ behavior is shown in figure 10 for the following parameters: $K = 6\,\text{pN}$ [32], $\sqrt{2}\pi\sigma_0\left(\sqrt{2} - 2 - w\right) = -10^{-5}\,\text{J/m}^2$, $\sqrt{A/\pi} = 5\,\mu\text{m}$, $L = 100\,\mu\text{m}$; according to this dependency, the single core will be expected to split into two cores at about $30.3^\circ\text{C}$. Similar consideration can be applied to $m = -1$ disclination splitting into $m = -1/2$ defects, figure 9.

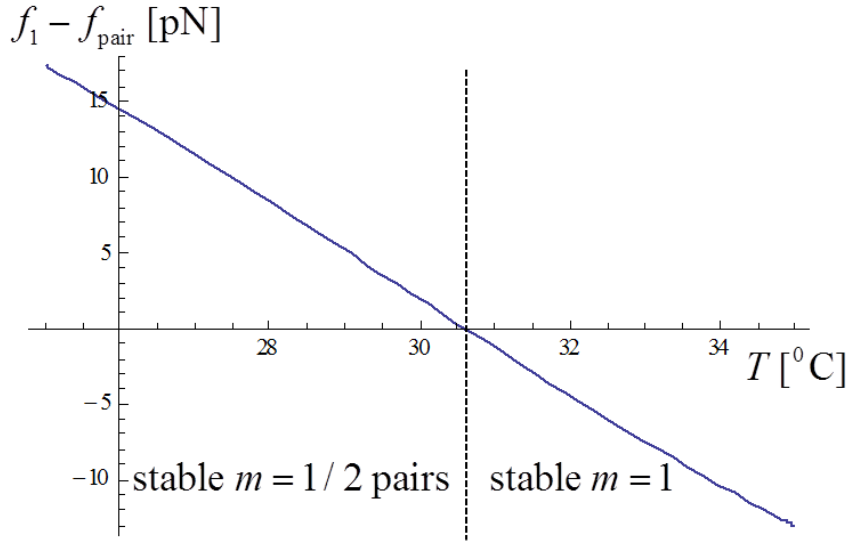

**Figure 10.** Temperature dependence of the free energy difference between the single-core $m = 1$ disclination and a pair of $m = 1/2$ disclinations.

The qualitative features of the model above, such as relative stability of integer defects with a large isotropic core at high temperatures, their splitting into pairs at low temperatures and separation of the semi-integer defects of the same sign are clearly observed in the experiments, figure 8, 9. For a quantitative comparison, much more information on the materials parameters needs to be gathered. Note that in the homogeneous N phase, we observe only semi-integer disclinations that we discuss in a greater detail below.

**8. Homogeneous N; cores of semi-integer disclinations.**

Since the director anchoring is degenerate in the plane of glass plates, the N textures typically contain a certain number of semi-integer disclinations, figure 11(a) and 12(a). Even in the deep N phase, well below the temperature range of the biphasic region, the cores of disclinations in LCLC show unusual and interesting features. First, the cores are not circular and feature one (in case of $m=1/2$) or three (in case of $m=-1/2$) cusp-like irregularities, located in the region where the director is forced to be parallel to the radius-vector emanating from the geometrical center of the defect. Second, the distance over which the structure shows a substantial change in the degree of orientational order, is macroscopic, on the order of $10\,\mu\text{m}$. In Figure 11(b,c), we show the optical retardance $\Gamma = |n_e - n_0|d$ profile measured across the disclination $m=1/2$ core. Far away from the core, the retardance is practically constant. If the temperature is raised, the far-field retardance slightly decreases, as the LCLC aggregates become shorter so that birefringence $|n_e - n_0|$ and scalar order parameter $S$ are reduced. Within the core region of the linear size of about $10\,\mu\text{m}$, the retardance drops sharply to zero, being practically a linear function of the distance to the center at which $\Gamma(r=0)=0$. Below we discuss the possible origin of cusps, the linear $\Gamma(r)$ dependence and the large core size of disclinations in the deep N phase of LCLC.

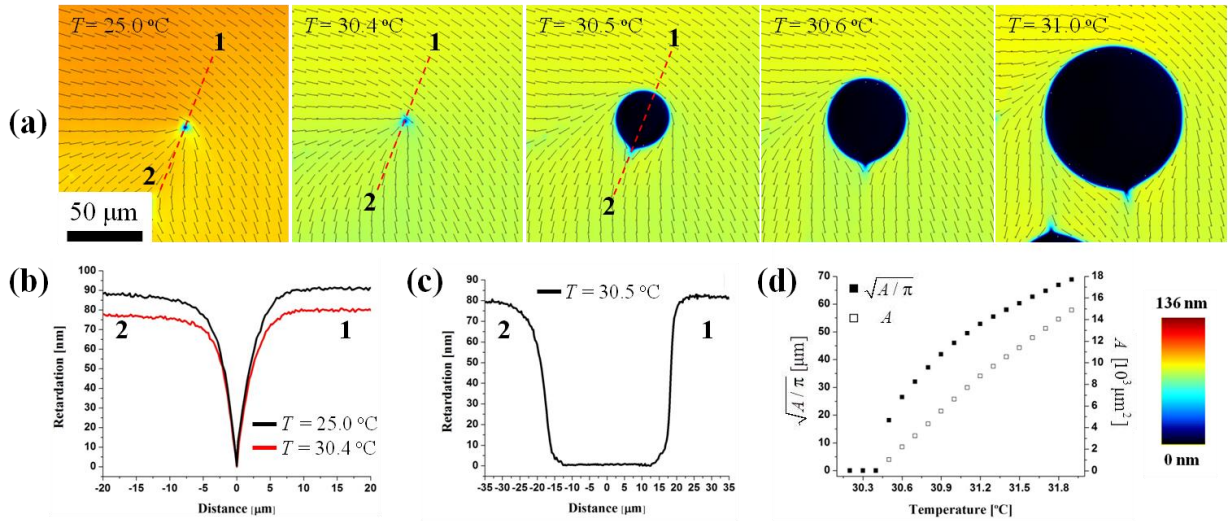

**Figure 11.** (a) PolScope textures of $m=1/2$ disclination and a 1c negative tactoid at the core of the disclination, as a function of increasing temperature. (b) Optical retardance $\Gamma$ across the core of defect shown in part (a) for 25 $^0$C and 30.4 $^0$C. (c) Optical retardance $\Gamma$ across the expanded I core of the defect in the biphasic region, corresponding to the temperature 30.5 $^0$C. (d) Temperature dependence of characteristic size and area of the 1c negative tactoid.

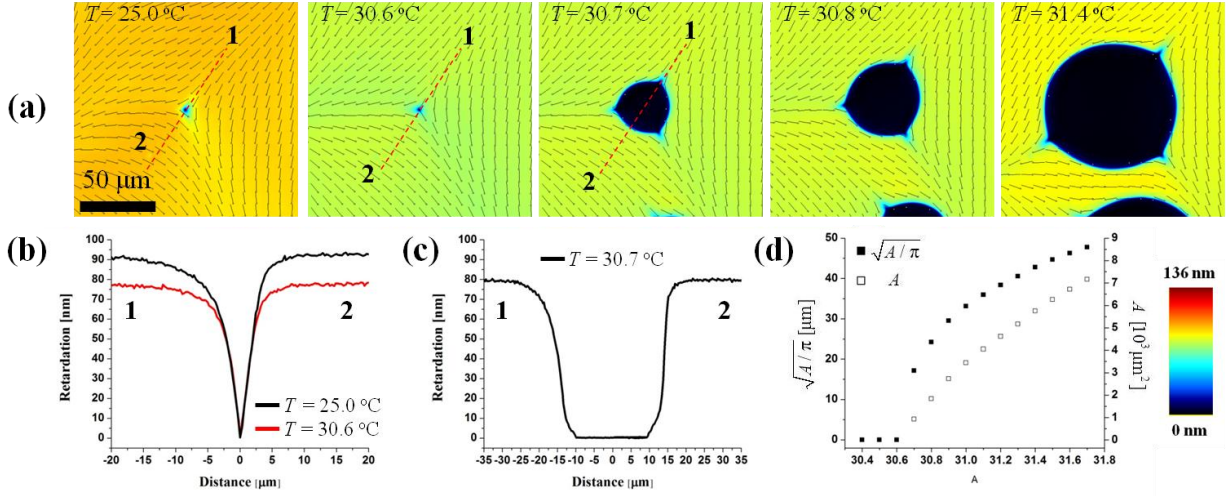

**Figure 12.** (a) PolScope textures of $m = -1/2$ disclination and a 3c negative tactoid at the core of the disclination, as a function of increasing temperature. (b) Optical retardance $\Gamma$ across the core of defect shown in part (a) for 25 $^0$C and 30.6 $^0$C. (c) Optical retardance $\Gamma$ across the expanded I core of the defect in the biphasic region, corresponding to the temperature 30.7 $^0$C. (d) Temperature dependence of characteristic size and area of the 3c negative tactoid.

The nematic orientational order is described by a traceless symmetric tensor order parameter [10]

$$Q_{ij} = S\left(\hat{n}_i\hat{n}_j - \delta_{ij}/3\right) + P\left(\hat{l}_i\hat{l}_j - \hat{m}_i\hat{m}_j\right) \tag{8}$$

which is diagonal $\mathbf{Q} = \mathrm{diag}(P - S/3, -P - S/3, 2S/3)$ in the frame of three mutually orthogonal directors $\hat{\mathbf{l}} \equiv -\hat{\mathbf{l}}$, $\hat{\mathbf{m}} \equiv -\hat{\mathbf{m}}$, and $\hat{\mathbf{n}} \equiv -\hat{\mathbf{n}}$; the quantities $S$ and $P$ are the uniaxial and biaxial order parameters, respectively, that depend on temperature, pressure and composition. In this work, we are dealing with a uniaxial N phase, thus $P = 0$, at least in the uniform state. The free energy functional associated with the nematic order can be written in the Landau- de Gennes form as [10]

$$f_0 = \tfrac{1}{2}a(T - T^*)Q_{ij}Q_{ji} - \tfrac{1}{3}BQ_{ij}Q_{jk}Q_{ki} + \tfrac{1}{4}C\left(Q_{ij}Q_{ji}\right)^2, \tag{9}$$

where $a$, $T^*$, $B$, and $C$ are material parameters.

The presence of the cubic term is significant, as it underscores the first order character of the I-N transition. Experimentally, this transition is weak. The experimental values for the coefficients above have been measured for thermotropic LCs. For example, for 5 CB, the often cited [73] values are $a = 0.13 \times 10^6$ J/$(\mathrm{m}^3 \cdot \mathrm{K})$, $B = 1.6 \times 10^6$ J/m$^3$, $C = 3.9 \times 10^6$ J/m$^3$. The order parameter can be spatially nonuniform, which is obviously the case of the first-order phase transitions in which the N and I phases can coexist and there is thus an interface that separates the two phases. The Landau-de Gennes expansion above is supplemented with the gradients of the tensor $\mathbf{Q}$ that in the second order expansion reads:

$$f_{LdG} = f_0 + L_1 Q_{ij,k} Q_{ij,k} + L_2 Q_{ij,j} Q_{ik,k} + L_3 Q_{ij,k} Q_{ik,j}, \tag{10}$$

where $L_n$ are the elastic constants and $_{,k} = \partial/\partial x_k$ are the spatial derivatives. It is convenient to rewrite $f_{LdG}$ in the form

$$f_{LdG} = f_0 + L_1 Q_{ij,k} Q_{ij,k} + L_a \left( Q_{ij,j} Q_{ik,k} + Q_{ij,k} Q_{ik,j} \right) + L_s \left( Q_{ij,j} Q_{ik} - Q_{ij} Q_{ik,j} \right)_{,k}, \tag{11}$$

that singles out the divergence (last) term; here $L_a = (L_2 + L_3)/2$ and $L_s = (L_2 - L_3)/2$. This form is useful for description of the I-N interface and defect cores, because its structure is represented by a continuous $\mathbf{Q}$-tensor field and the divergence term becomes negligible. When the $\mathbf{Q}$-tensor changes over the distances much larger than the molecular scale and the $\mathbf{Q}$-tensor gradients are weak (e.g. far away from the defect core), the scalar order parameters adopt constant values that correspond to the minimum of $f_0$, $S = S_N$, $P = 0$. Then the elastic energy density is controlled by the director distortions as described by the Frank-Oseen functional:

$$f_{FO} = \tfrac{1}{2} K_1 (\operatorname{div} \hat{\mathbf{n}})^2 + \tfrac{1}{2} K_2 (\hat{\mathbf{n}} \cdot \operatorname{curl} \hat{\mathbf{n}})^2 + \tfrac{1}{2} K_3 (\hat{\mathbf{n}} \times \operatorname{curl} \hat{\mathbf{n}})^2 - K_{24} \operatorname{div} (\hat{\mathbf{n}} \operatorname{div} \hat{\mathbf{n}} + \hat{\mathbf{n}} \times \operatorname{curl} \hat{\mathbf{n}}), \tag{12}$$

with the splay, twist, bend and saddle-splay terms, respectively.

Comparison of $f_{LdG}$ and $f_{FO}$ results in the following relations between the elastic constants $K_1 = K_3 = 4 S_N^2 (L_1 + L_a)$, $K_2 = 4 S_N^2 L_1$, and $K_{24} = S_N^2 (2 L_1 + L_a - L_s)$; the difference between splay and bend constants starts with the cubic term $K_1 - K_3 = O(S_N^3)$ [74-77], $L_1$ corresponds to the one elastic constant approximation, and $L_a$ provides elastic anisotropy. Below we investigate the effect of $L_a$ on the N-I interface and the defect cores.

Consider a planar N-I interface normal to the $x$-axis at the transition temperature $T_{NI}$ when $f_0(S_N) = 0$. The energy of the N-I interface per unit area can be calculated by minimizing $F = \int f_{LdG} dx$, where $f_{LdG}$ is the characteristic width of the N-I interface [74]. Neglecting the possible biaxiality and assuming that the director does not reorient across the interface, one obtains

$$F = 2 a T_{NI} h \int_0^{S_N} \sqrt{f_0(S)} \, dS, \tag{13}$$

where the coefficient $h$ takes a different value for the director normal to the interface, $h = h_n = \sqrt{(3 L_1 + 4 L_a)/a T_{NI}}$, and for the tangential director, $h = h_t = \sqrt{(3 L_1 + L_a)/a T_{NI}}$. The dependence of the interface energy on the director orientation allows us to estimate the surface anchoring coefficient $w = h_n/h_t - 1 = \sqrt{(3 L_1 + 4 L_a)/(3 L_1 + L_a)} - 1 \approx \sqrt{(4 K_1 - K_2)/(K_1 + 2 K_2)} - 1$. Consider now the disclination core.

The $\mathbf{Q}$-tensor field of the disclination core is determined by the functional $F = \int f_{LdG} dV$. We consider wedge planar disclinations with a two-dimensional director field $\hat{\mathbf{n}}$ independent of the $z$-coordinate along the constant $\hat{\mathbf{m}}$ director normal to the bounding plates. We describe the $\mathbf{Q}$-tensor, equation (8), in the polar coordinates $(r, \theta)$ with the disclination center at $r = 0$ by order parameters $S(r, \theta)$, $P(r, \theta)$ and the angle $\varphi(r, \theta) = m\theta + \tilde{\varphi}(r, \theta)$ between $\hat{\mathbf{n}}$ and $x$-axis; $\tilde{\varphi}(r, \theta)$ is a correction term. Mottram and Hogan [78] assumed that the core is axially symmetric, so that $S(r)$, $P(r)$ and $\varphi(r)$ are $\theta$-independent. This assumption is correct if $m = 1$ or when the Frank constants are equal, $L_a = 0$. For disclinations with $m \neq 1$ the minimum of energy is reached by including angular harmonics $k(m-1)\theta$, e.g. $S(r,\theta) = S_0(r) + \sum_{k=1} S_k(r)\cos[k(m-1)\theta]$. In LCLCs, a strong anisotropy $L_a \neq 0$ of elastic constants has been demonstrated, with $K_2 \ll K_1, K_3$ [32]. Therefore, we expect $w \sim 1$, which leads to a structure that is not axially symmetric and features $2|m-1|$ cusps. The largest asymmetry is observed for $m=1/2$, as illustrated by the spatial profiles of scaled order parameters $\hat{S} = (C/B)S$, $\hat{P} = (C/B)P$ in figure 13(a) and of optical retardance $\hat{\Gamma} = \hat{S} - \hat{P}$ in figure 13(b). The simulated retardance $\hat{\Gamma}$ is very close to the experimental retardance profile, see figure 11(b). The only difference is that the experimental core does not change as much with temperature (within the temperature range of the N phase) as the theory predicts, figure 13(a). The most probable reason is the balance of surface tensions between the I phase, N phase and the substrates at the bounding plates that are not accounted for in the model.

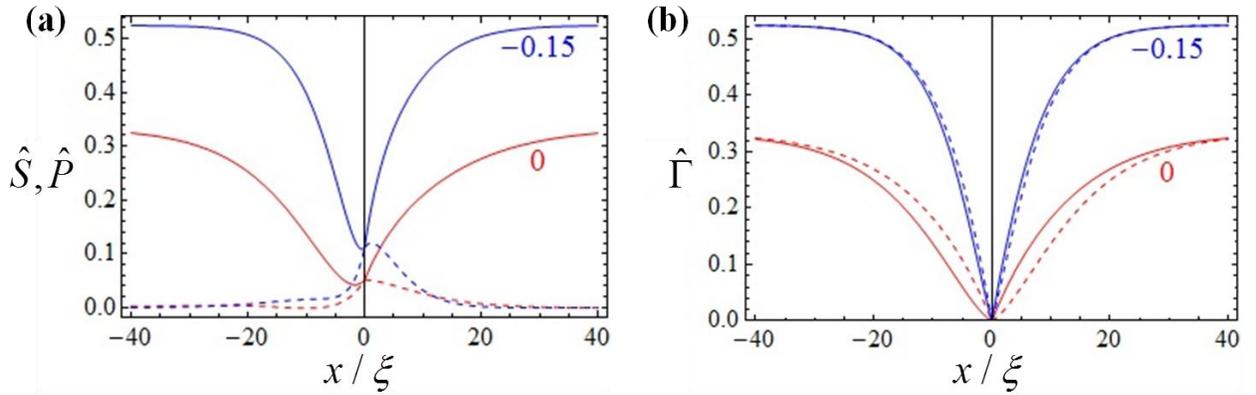

**Figure 13.** Simulations of core of $m=1/2$ disclination: (a) dependence of order parameters $\hat{S}$ (solid) and $\hat{P}$ (dashed) on the dimensionless distance $x/\xi$, $\xi = \sqrt{L_1 C}/B$, along the symmetry line at the temperatures $a(T-T^*)C/B^2 = 0$ (red) and $= -0.15$ (blue); (b) the normalized phase retardances $\hat{\Gamma} = \hat{S} - \hat{P}$ (solid lines) calculated along the symmetry line at the same temperatures; dashed curves are mirror images of the solid curves and demonstrate the structural asymmetry caused by a relatively large elastic anisotropy $L_a/L_1 = 4$.

The biaxial nature of the $\pm 1/2$ disclinations' cores was discussed by Lyuksyutov [79]: as one approaches the center of the defect, $r \to 0$, the energy density of the director distortions $K/r^2$ becomes comparable to the cubic term $BS^3$ in the Landau- de Gennes free energy expansion. The presence of the cubic term is significant, as it underscores the first order character of the I-N transition. At the scales $r_B \sim \sqrt{\dfrac{K}{BS^3}}$ and smaller, the distortions can be relaxed by allowing a structure to become biaxial, as described in details by Schopol and Sluckin [80]. In thermotropic 5CB, with the typical estimates $K = 10\,\text{pN}$, $B = 1.6 \times 10^6\,\text{J/m}^3$, $S = 0.6$ [80], one finds $r_B \sim 10\,\text{nm}$, a very small (few molecular scales) value, hardly accessible to the modern imaging techniques. This is why the theoretical models of Lyuksyutov [79] and Schopol-Sluckin [80] have not been verified so far. Since the core size is inversely proportional to the coefficient $B$, it should become large as one approaches a point at the phase diagram where $B$ vanishes. In our case, the scalar order parameter changes over the scale of 10 μm. Since the elastic constants and $S$ in LCLC nematics are of the same order of magnitude as their counterparts in thermotropic LCs such as 5CB, the observation of a huge radius of disclination $r_B \sim 10\,\mu m$, figures 11(b) and 12(b), suggests that in DSCG, the coefficient $B$ might be extremely small, only about $B = (0.1 - 1)\,J/m^3$. Interestingly, Tang et al [81, 82] reported that in a lyotropic N formed by filaments of f-actin, the I-N transition is continuous when the average length of the filaments is longer than 2 μm and is of the first order if the filaments are shorter. Another mechanism of the enlarged core might be brought about by the aggregate structure of LCLCs. At the core, the average length of the aggregates might be smaller than their length in a uniform N, thus the set of parameters in the Landau-de Gennes expansion becomes a function of position in space near the defect core [83].

### 9. N-to-I transition: I tactoids and multiply connected tactoids

When the temperature is increased to some critical value, the cores of disclinations start to expand dramatically, see figure 11 for $m = 1/2$ and figure 12 for $m = -1/2$. To characterize the temperature behavior of the cores, we use the normalized square root $\sqrt{A/\pi}$ of the isotropic (black in PolScope textures) area $A$ as a measure of core size. As seen from data in figure 11(d) and 12(d), the temperature dependence of the core size for $m = \pm 1/2$ defects is close to $\sqrt{A} \propto const + \sqrt{T - T_c}$, where $T_c$ is some critical point at which the core starts its expansion; it slightly varies from defect to defect (even if they are of the same charge), indicating an influence of local effects such as surface irregularities. Defect-free uniform portions of the N phase also nucleate I islands, in the shape of negative tactoids with two cusps, figure 14. The temperature dependence of the negative 2c tactoid size, also measured as $\sqrt{A/\pi}$, figure 14(b), is very similar to the behavior observed for the isotropic disclination cores, figures 11(d) and 12(d).

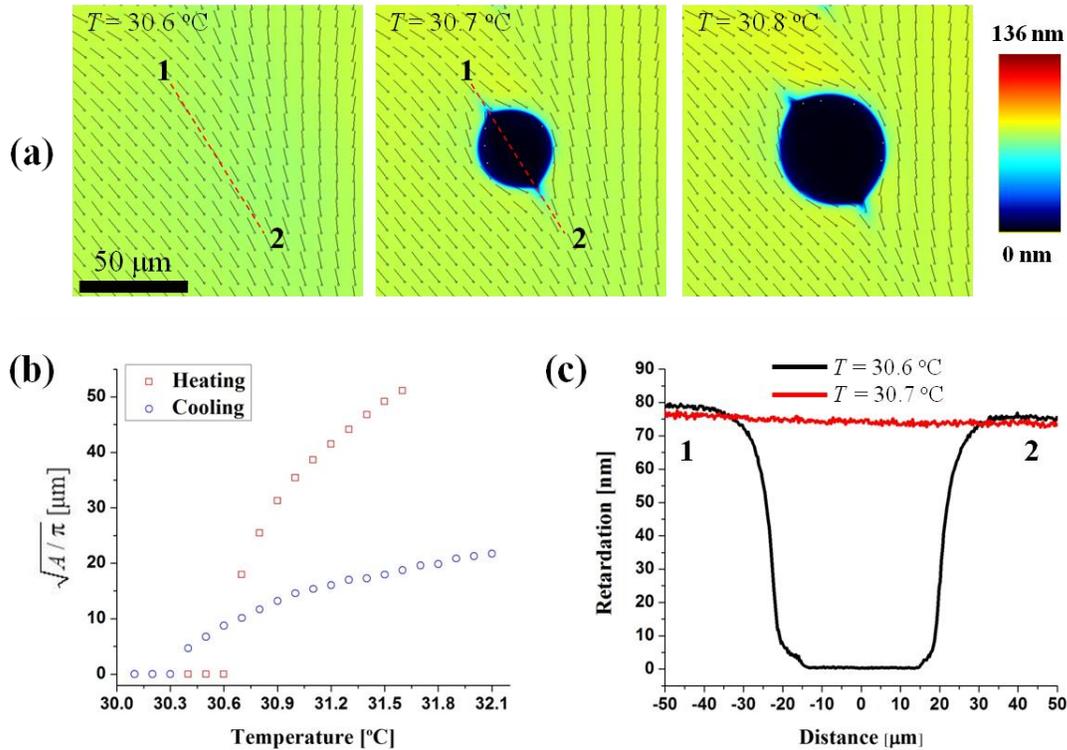

**Figure 14.** (a) PolScope textures of a 2c negative tactoid nucleated in the uniform N part of texture. (b) Temperature dependence of characteristic size and area of the 2c negative tactoid. (c) Optical retardance $\Gamma$ before and after the appearance of the 2c tactoid.

The temperature dependence of the core size has been analyzed by Mottram and Sluckin [84] for $m = 1/2$, under an assumption that the core represents a round island of an I phase. The disclination energy is then comprised of the elastic energy of the director distortions outside the core, surface energy of the N-I interface $2\pi\sigma_0 r_{1/2}$, and the condensation energy $(f_I - f_N)\pi r_{1/2}^2 \propto (S_I - S_N)(T_{NI} - T)r_{1/2}^2$, where $S_I$ and $S_N$ are the entropy densities of the I and N phase, respectively. The core size was shown to grow with temperature, following the parametric dependence $T(r_{1/2}) = T_{NI} + \frac{\sigma_0}{(S_I - S_N)r_{1/2}} - \frac{K}{8(S_I - S_N)r_{1/2}^2}$ with $\partial r_{1/2}/\partial T > 0$ and $\partial^2 r_{1/2}/\partial T^2 > 0$.

In the case of the biphasic region of LCLC, the core size of disclinations does not follow this model, since experimentally, $r_{\pm 1/2} \propto const + \sqrt{T - T_c}$, so that $\partial r_{\pm 1/2}/\partial T > 0$ but $\partial^2 r_{\pm 1/2}/\partial T^2 < 0$. The distinctive behavior of the core defects in LCLCs is brought about by the broad temperature range of I-N coexistence in which the total area (volume) of each of the two phases shows an approximately linear dependence on $T$, figure 2(d). The total energy of a singular $m = \pm 1/2$ disclination can be approximated as $f_{\pm 1/2} = f_{e,\pm 1/2} + f_{cond,\pm 1/2} + f_{IN,\pm 1/2} = \frac{\pi}{4} K \ln R/r_{\pm 1/2} + \pi r_{\pm 1/2}^2 (f_I - f_N) + 2\pi r_{\pm 1/2} \sigma_{r_{\pm 1/2}}$, where the equilibrium core size $r_{\pm 1/2} = \frac{1}{f_I - f_N}\left[\sqrt{\sigma_{\pm 1/2}^2 + K(f_I - f_N)/2} - \sigma_{\pm 1/2}\right]$ is determined by the anisotropic surface energy $\sigma_{\pm 1/2}(\sigma_0, w, m)$ that is the function of the isotropic surface tension, surface anchoring strength $w$ and the strength of disclination.

The negative tactoids contain negative cusps, i.e. regions of the I phase protruding into the N phase at the points where the director pattern of the disclination forces the director to become perpendicular to the I-N interface. In principle, the LC can simply reduce its scalar order parameter in these regions to reduce the energy of director distortions or realign the director along the normal to the bounding plates, figure 11, 12. As already indicated in the discussion of the splitting of integer disclination, the second mechanism is less likely. In the N regions around the negative cusps, one finds negative boojums with $m = -1/2 - \tau/2\pi$. Similarly to the conservation law (4), one can write down a conservation law for the negative tactoid surrounded by the director field that corresponds to a disclination with the strength $m_{out}$:

$$\sum_i^c \left(m_i + \frac{\tau_i}{2\pi}\right) - m_{out} = -1, \tag{14}$$

which leads to the relationship between the number of positive and negative cusps and $m_{out}$:

$$c^- - c^+ = 2(1 - m_{out}). \tag{15}$$

For a uniform far field, $m_{out} = 0$, there are two more negative cusps than the positive ones, for $m_{out} = -1/2$, there should be at least three negative cusps, etc.

Comparison of figures 11, 12 and 14 demonstrates that the 1c, 3c and 2c isotropic tactoids occur in a close proximity to each other on the temperature scale, but still at somewhat different values of $T_c$. The cores of the disclinations are not necessarily the ones that start to expand first; sometimes the 2c isotropic tactoids nucleate at temperatures lower than the temperature at which the disclination core already present in the system starts to expand. The overall expansion of the isotropic tactoids with temperature follows the temperature behavior of the isotropic phase fraction shown in figure 2(d).

The dynamics of the phase transition can be used to create I-N-I tactoids that are non-simply connected. These represent N tactoids that contain large I tactoids inside them, figure 15; "large" here means larger than the disclination core in a homogeneous N phase. A practical approach to create I-N-I tactoids is to nucleate first N tactoids by cooling the system from the I phase and then rapidly heat it up, remaining within the biphasic region.

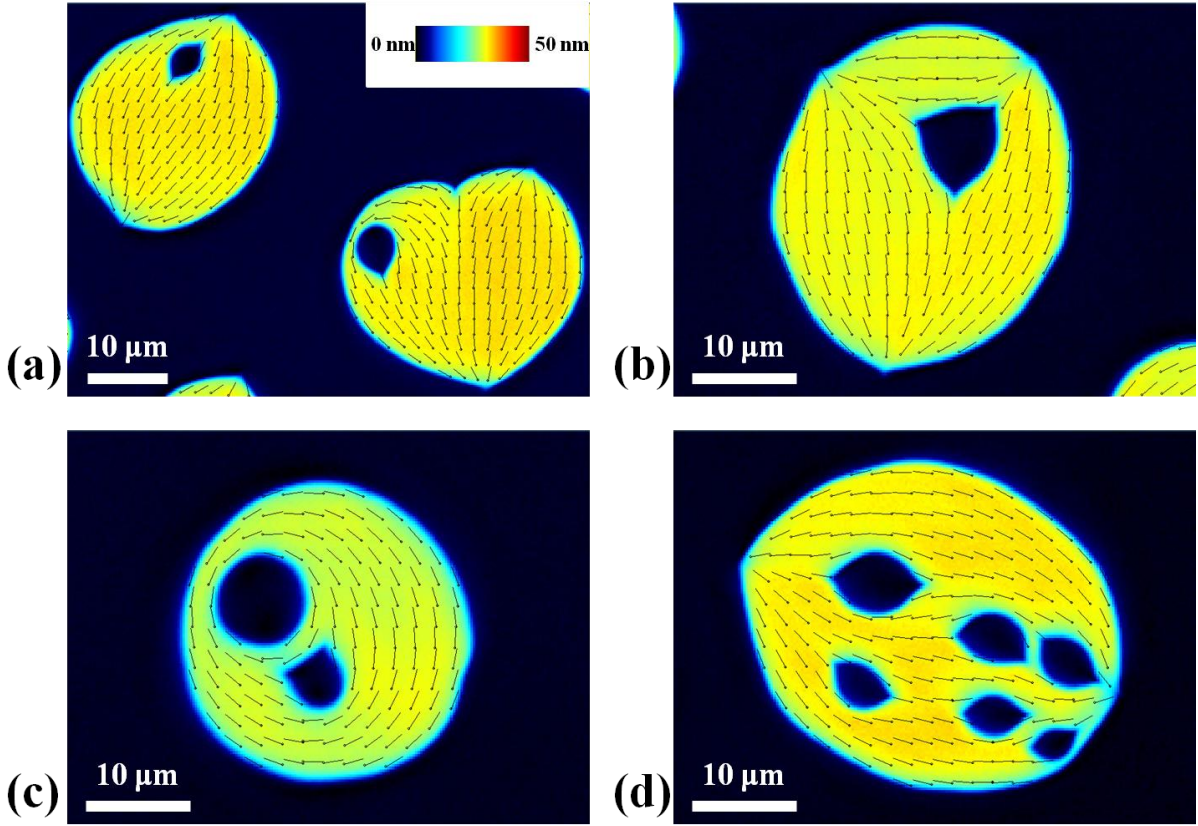

**Figure 15.** PolScope textures of I-N-I tactoid-in-tactoid multiply connected forms with different number of the isotropic inclusions: (a) $I=1, c^-=c^+=2$, (b) $I=1, c^-=c^+=3$, (c) $I=2, c^-=2, c^+=0$, and (d) $I=6, c^-=12, c^+=2$.

The conservation law for these more complicated geometries in Fig. 15 writes as a generalization of Eqs. (4) and (14):

$$\sum_i^c \left(m_i + \frac{\tau_i}{2\pi}\right) + \sum_j^n m_j = 1 - I \quad \text{or} \quad \frac{1}{2}\sum_i^c k_i + \sum_j^n m_j = 1 - I, \tag{16}$$

where $c$ is the number of boojums at all the I-N interfaces, $m_i$ is their individual strength, $\tau_i$ are the cusp angles corresponding to the $i$-th boojum, $n$ is the number of disclinations in the N interior of the I-N-I tactoid, with topological strengths $m_j$, and $I$ is the number of isotropic tactoids inside the nematic tactoid. Similarly to the previously described cases of simply-connected N and I tactoids, the relationship (16) can be rewritten in terms of the excess number of positive cusps $c^+$ over the negative ones $c^-$ residing at all the I-N interfaces of the multiply connected I-N-I tactoid:

$$c^+ - c^- = 2\left(1 - I - \sum_{j}^{n} m_j\right). \quad (17)$$

As an illustration, figure 15(a) shows two I-N-I tactoids with $I = 1$ and two different realizations of the condition $c^+ - c^- = 0$. Figure 15(b) illustrates the case of $c^+ = c^- = 3$. Figure 15(c) shows an I-N-I tactoid with two I regions. In this particular case, there are no cusps on two round I-N interfaces; the only cusps are two negative cusps at the interior 2c I tactoid. Finally, figure 15(d) shows six 2c I tactoids inside one N tactoid, $c^+ - c^- = -10$.

## 10. Shape of N and I tactoids in frozen director field.

As already indicated, the shape of the positive and negative tactoids is determined by the balance of the anisotropic surface tension and nematic elasticity of the region either inside the positive tactoid or outside the negative tactoid. There is no analytical solution of this problem. On the other hand, the LCLC tactoids observed in this work show some general features such as multiple cusps, the number of which is determined by the topological structure of associated disclinations. Therefore, it is of interest to describe these forms using the simplest possible approximation, namely, by assuming that the director field is frozen in its distorted state, as determined by the disclination structures written in polar coordinates ($r, \theta$) as $n_x = \cos(m\theta + const)$, $n_y = \sin(m\theta + const)$; examples of different $m$'s are shown in figures 4(a-d). This approximation is equivalent to an assumption of an infinitely large elastic constants of the N phase. With the director field frozen, the shape of the I-N boundary of the tactoids is determined by the angular dependence of the surface tension that we take in the form $\sigma(\alpha) = \sigma_0\left(1 + w\cos^2\alpha\right)$. We place the geometrical center of the tactoid (positive or negative) at the center of disclination. The angular dependence of the surface energy at the I-N interface drawn in the frozen director field then becomes dependent on the disclination configuration $(m)$ and leads

$$\sigma(\theta) = \sigma_0\left\{1 + w\cos^2\left[(m-1)\theta + const\right]\right\}, \quad (18)$$

where the constant in the argument specifies the overall orientation of the director pattern; it can be put equal 0.

Applying the principle of Wulff construction of equilibrium shapes of perfect crystals [68, 85-89] to the disclination-specific expression (18) for the anisotropic interfacial tension, we construct the equilibrium shapes of negative and positive 2D tactoids under the condition of a constant surface area and a frozen director field. The principle is illustrated in figure 16.

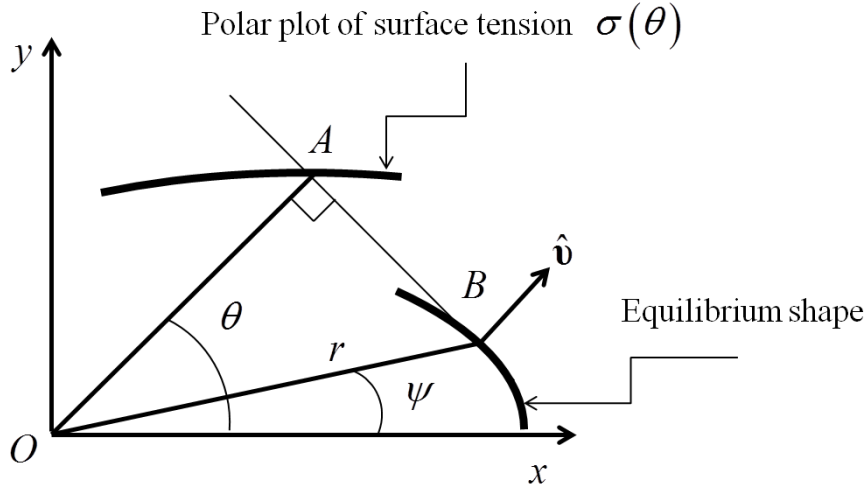

**Figure 16.** Scheme of Wulff construction of an equilibrium shape of a tactoid of a constant area from the polar plot of the surface tension.

From some origin $O$, we draw a line segment $OA$ along the direction $\theta$, of a length $\sigma(\theta)$. Repeating for all angles $\theta$, we obtain the polar plot of anisotropic tension $\sigma(\theta)$. In each point $A$ of $\sigma(\theta)$, raise the normal to the segment $OA$. The equilibrium shape is the pedal of the polar plot $\sigma(\theta)$, i.e., the convex envelope of all these normal. The coordinates $(x, y)$ of the shape curve are expressed in terms of the polar angle $\psi$ between the normal to the I-N interface and the $x$-axis and the radius-vector $r = OB$. As easy to see from figure 16, $r\cos(\psi - \theta) = \sigma(\theta)$. By taking the derivative with respect to $\theta$, $r\sin(\psi - \theta) = \sigma'(\theta)$, one obtains two equations that provide a parametric description of the tactoid's shape: $r = \sqrt{\sigma^2 + \sigma'^2}$ and $\psi = \theta + \arctan\dfrac{\sigma'}{\sigma}$. In the Cartesian coordinates, the shape is parametrized as

$$x(\psi) = \sigma(\psi)\cos\psi - \sigma'(\psi)\sin\psi;$$
$$y(\psi) = \sigma(\psi)\sin\psi + \sigma'(\psi)\cos\psi. \qquad (19)$$

Figure 17(a) shows different polar plots of surface potential $\sigma(\psi)$ with $w = 2$ for different director configurations, shown in figure 4(a)-(d), as well as for the uniform director field, $m = 0$. The corresponding shapes of tactoids are shown in figure 17(b). These shapes should be understood as the shape separating either the I interior from the N exterior (negative tactoids) or the N interior from the I exterior (positive tactoids).

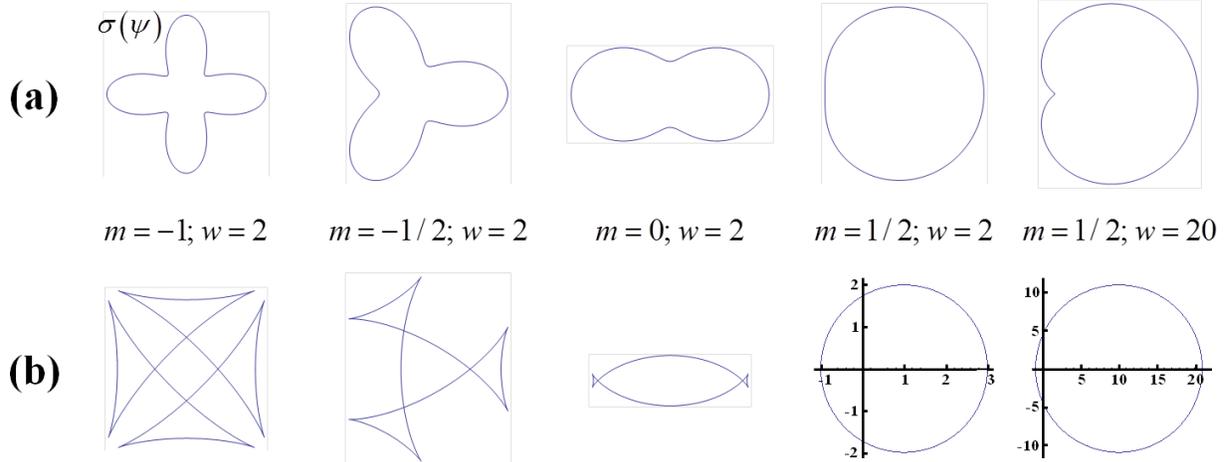

**Figure 17.** (a) Polar plots of anisotropic surface tension, equation (18), corresponding to the frozen director fields with different topological charges *m* shown in figures 4(a-d). (b) The corresponding equilibrium shapes of the negative and positive tactoids, equations (19). The physically relevant parts of the shape do not include the triangular "ears" attached to the cusp points.

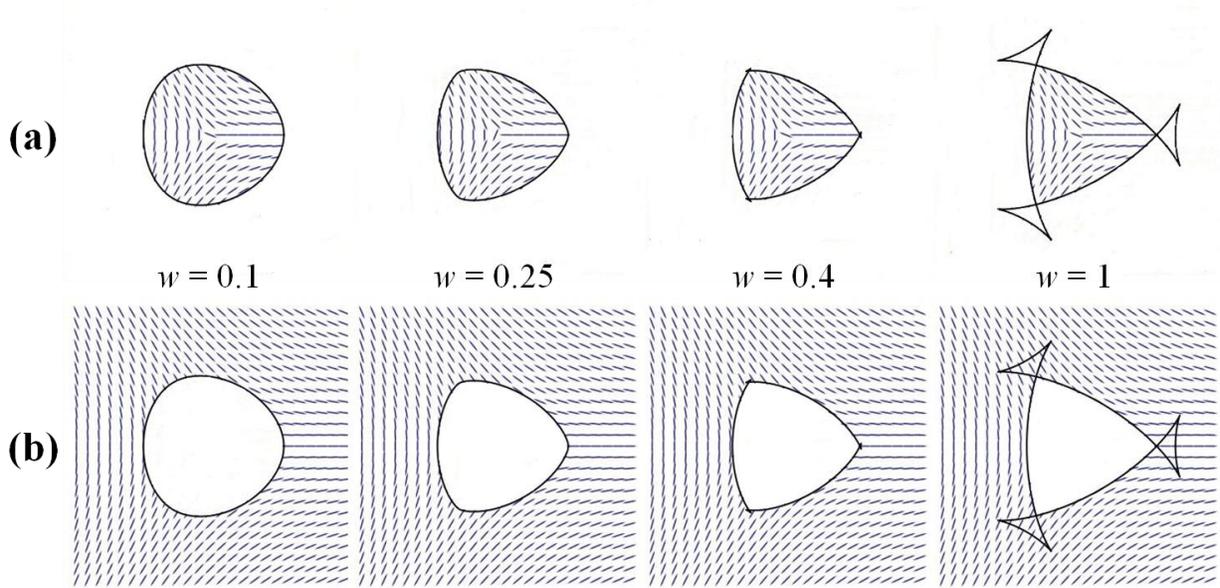

**Figure 18.** Illustration of how the surface anchoring shapes up the (a) positive N and (b) negative I tactoids with a frozen director of strength $m = -1/2$.

Three configurations, namely, $m = -1; -1/2; 0$ for a chosen value of surface anisotropy $w = 2$ show shape contour that crosses itself, figure 17(b). These crossings are associated with zeroes of the radius of curvature of the envelope, which is $R = \sqrt{r'^2 + r^2 \psi'^2} = \sigma + \sigma''$ [68, 87-89]. Whenever $\sigma + \sigma'' > 0$, the resulting shape is rounded, with the I-N interface showing all possible orientations in the plane. If $\sigma + \sigma'' < 0$, the pedal develops cusps of angular discontinuity: some of the orientations of the I-N interface are missing from the equilibrium shape. The criterion $\sigma + \sigma'' < 0$ for missing orientations for a tactoid associated with the director field of strength $m$ and the surface tension dependency given by equation (18), reads

$$1 + w\left[1 - 4(m-1)^2\right]\cos^2(m-1)\psi + 2(m-1)^2 w < 0. \tag{20}$$

For $m = 1/2, 1$, this condition is never fulfilled and the shape of the tactoids is circular. Interestingly, as the surface anchoring increases, the core of the $m = 1/2$ disclination shifts towards the axis at which the director is normal to the I-N interface, see figure 17(b); in this particular model, however, the core remains circular. For all other director configurations, the center of the core coincides with the center of the director field.

For $m = -1/2$ configurations, inequality (20) predicts $\sigma + \sigma'' < 0$ when $w > 2/7$. Figure 18 shows a progressive change in the shape of 3c tactoid with $m = -1/2$ as the surface anisotropy increases from the values below the critical value $w_{-1/2} = 2/7$ ($w = 0.1; 0.25$) to the values above $w_{-1/2}$ ($w = 0.4; 1$). When $w < w_{-1/2}$, the I-N interface is round with all possible orientations. When $w > w_{-1/2}$, some orientations are missing. The geometrical indicator of missing orientations are triangular "ears" attached to the cusps in which $\hat{\mathbf{n}}$ is perpendicular to the interface. These ears do not correspond to physical portions of the shape. Our analysis is equally applicable to the positive 3c tactoids shown in figure 18(a) and to the negative 3c tactoids shown in figure 18(b). For $m = -1$, the condition of missing orientations of the I-N interface reads $w > w_{-1} = 1/7$ and for a classic 2c tactoid with $m = 0$, $w > w_0 = 1$.

## 11. CONCLUSION

The term morphogenesis is usually reserved for the development of forms and structures in systems of biological origin. In this work, we illustrate how the balance of anisotropic surface energy and internal elasticity shapes up complex morphogenetic developments of tactoidal forms and topological defects emerging in the dynamic phase transitions between the isotropic and nematic phases. In the simplest case of a thermotropic liquid crystal, the balance of the surface and bulk effects produces an equilibrium set of topological defects in N droplets that are larger than about 1-10 micrometers. The shape of these N droplets is close to spherical. The situation is very different in the lyotropic chromonic liquid crystal DSCG that was studied in details for a 2D geometry of confinement. The nuclei of new phase appear as strongly anisometric islands with pronounced cusps at which the I-N interface shows a singular behavior and the director forms surface defects-boojums. The reason for the existence of defects is the anisotropy of interfacial tension, i.e., surface anchoring of the director. The growing nuclei of the N phase merge with each other and often create bulk defects-disclinations of semi-integer or even integer strength in a process similar to the Kibble mechanism. The presence of interior disclinations influences the number of cusps featured by the closed I-N interface. We derive conservation laws that relate the number of positive and negative cusps $c$ to the topological strength $m$ of the defects inside and outside the simply-connected I and N tactoids. Similar relationships are established for the experimentally observed multiply connected tactoids.

The integer-strength disclinations that might exist in the biphasic region, split into pairs of semi-integer disclinations upon approaching the homogeneous N phase. In the homogeneous N phase, well below the temperatures of biphasic region, only semi-integer disclinations survive. The cores of these disclinations are of an extraordinary large radius (tens of microns) and of a non-circular shape. Using the Landau-de Gennes theory, we demonstrate that the disclination cores contain cusp-like regions, as in the experiment. The theory associates the cusps at the disclination cores with the anisotropy of elastic constants that translates into an effective anisotropy of the I-N interface. In the so-called one-constant approximation, the model predicts cusps-free circular cores. The N-to-I transition features isotropic 2c tactoids nucleating in the uniform parts of the texture and tactoids with one or three cusps nucleating at the cores of $m = 1/2$ and $m = -1/2$ disclinations, respectively. The size of these isotropic cores increases with temperature, approximately as $r_{\pm 1/2} \propto const + \sqrt{T-T_c}$, which is consistent with the overall temperature dependence of the volume fractions of coexisting I and N phases. Finally, we illustrate how the orientational anisotropy of interfacial tension leads to the formation of tactoidal shapes with cusps, by applying a generalized Wulff construction to the topologically nontrivial frozen director fields.

This is only the first attempt to describe, mostly qualitatively, a very complex morphogenesis of phase transition in LCLCs. Much more work is needed to achieve a better quantitative understanding. An ultimate goal would be to predict the shape and director configuration of tactoidal formations from the first principles, using the anisotropic surface energy and the elasticity of orientational order.

**Acknowledgments**

We thank Paul van der Schoot and Shuang Zhou for fruitful discussions, Luana Tortora and Chan Kim for the help with experiments. The work was supported by NSF DMR grants 1104850 and 11212878.